\documentclass[pre,onecolumn,showpacs,preprintnumbers,amsmath,amssymb,superscriptaddress]{revtex4-1}
\usepackage{graphicx}
\usepackage{dcolumn}
\usepackage{bm}
\usepackage[usenames]{color}

\usepackage[all]{xy}
\usepackage[toc]{appendix}
\usepackage{amsmath}
\usepackage{natbib}
\usepackage{longtable}
\def \bea{\begin{eqnarray}}
\def \eea{\end{eqnarray}}
\def \be{\begin{equation}}
\def \ee{\end{equation}}

\def\({\left(} \def\){\right)}
\DeclareMathOperator\arccoth{arccoth}
\begin{document}
\bibliographystyle{apsrev4-1}

\renewcommand{\figurename}{Figure}

\title{Long-range Acoustic Interactions in Insect Swarms: An Adaptive Gravity Model}
\author{Dan Gorbonos}
\affiliation{Department of Chemical Physics, The Weizmann Institute of Science, P.O. Box 26, Rehovot, Israel 76100}
\author{Reuven Ianconescu}
\affiliation{Department of Chemical Physics, The Weizmann Institute of Science, P.O. Box 26, Rehovot, Israel 76100}
\affiliation{Shenkar College of Engineering and Design, Ramat-Gan, Israel}
\author{James G. Puckett}
\affiliation{Department of Physics, Gettysburg College, Gettsyburg, Pennsylvania 17325, USA}
\author{Rui Ni}
\affiliation{Department of Mechanical and Nuclear Engineering, The Pennsylvania State University, University Park, Pennsylvania 16802, USA}
\author{Nicholas T. Ouellette}
\affiliation{Department of Civil and Environmental Engineering, Stanford University, Stanford, California 94305, USA}
\author{Nir S. Gov\footnote{Correspondence and requests for materials should be addressed to N.S.G. (email: nir.gov@weizmann.ac.il).}}
\affiliation{Department of Chemical Physics, The Weizmann Institute of Science, P.O. Box 26, Rehovot, Israel 76100}

\begin{abstract}
The collective motion of groups of animals emerges from the net effect of the interactions between individual members of the group. In many cases, such as birds, fish, or ungulates, these interactions are mediated by sensory stimuli that predominantly arise from nearby neighbors. But not all stimuli in animal groups are short range. Here, we consider mating swarms of midges, which interact primarily via long-range acoustic stimuli. We exploit the similarity in form between the decay of acoustic and gravitational sources to build a model for swarm behavior. By accounting for the adaptive nature of the midges' acoustic sensing, we show that our ``adaptive gravity'' model makes mean-field predictions that agree well with experimental observations of laboratory swarms. Our results highlight the role of sensory mechanisms and interaction range in collective animal behavior. The adaptive interactions that we present here open a new class of equations of motion, which may appear in other biological contexts.
\end{abstract}
\maketitle

\section{Introduction}

Collective behavior of groups of social animals is widespread in nature \cite{parrish1999}, and occurs on size scales ranging from single-celled organisms \cite{gerisch1982,drescher2009} to insects \cite{kelley2013,puckett2014,giardina2014plos,giardina2014prl} to larger animals such as birds \cite{ballerini2008b,nagy2010} or fish \cite{katz2011,herbert-read2011,berdahl2013}. Animals are thought to aggregate and move cooperatively for many reasons; collective behavior may, for example, reduce the risk of predation for an individual in a group \cite{hamilton1971,parrish1999}, promote efficient mating and decrease inbreeding in dispersed populations \cite{downes1969}, modulate the energetic cost of migration \cite{portugal2014}, or enable enhanced sensing \cite{berdahl2013}. Because of both its ubiquity and the potentially advantageous properties it conveys for groups, collective behavior has engaged a broad cross-section of scientists, ranging from physicists and applied mathematicians who hope to tease out the general principles that drive the emergence of collective states in non-equilibrium systems to engineers who hope to develop bio-inspired control strategies for distributed multi-agent systems.

Collective behavior therefore has a long modeling history \cite{parrish1999,okubo1986,bernoff2013nonlocal}. Models are useful both as a check on our fundamental understanding of the kinds of low-level interactions that lead to the emergent group properties and as a stepping stone to the design of engineered systems that exploit them. Many models treat the group as a collection of self-propelled agents that are in some way coupled \cite{vicsek1995,couzin2002}. Building such an agent-based model therefore requires several fundamental choices \cite{tempirical}. We must specify the base-case, non-interacting behavior of each individual; we must choose a functional form for the interactions that couple the individuals; we must decide whether these rules are uniform throughout the population and in time; and we must decide which individuals interact. Each of these choices can be difficult to make with certainty, and yet has significant ramifications for model performance and fidelity.

Here, we focus on the last of these modeling assumptions: the choice of which individuals interact. From passive observations alone, it is difficult to discern the correct interactions in a group of animals \cite{puckett2014}, since it requires the solution of a challenging inverse problem. Thus, it is common to replace the difficult-to-measure interaction network with the more straightforward proximity network \cite{bode2011}; that is, one assumes that the local neighborhood of an individual dominates that individual's behavior. This local neighborhood may be defined by, for example, either metric or topological distance \cite{ballerini2008,ginelli2010}, but it is still the local environment that is assumed to matter. This assumption is reasonable, and appears to be valid \cite{buhl2006,ballerini2008,lukeman2010,katz2011,herbert-read2011}, for dense groups of animals such as flocks of birds or schools of fish that interact primarily through vision and that move in a coordinated, directed fashion. But it need not always be true; crowds of humans moving toward goals, for example, can show emergent collective behavior while ``interacting'' with other individuals with whom they are likely to collide in the future rather than those who are closest to them \cite{guy2012}.

We consider here a canonical example of collective animal behavior---mating swarms of flying insects---where local interactions do not clearly play a major role and yet where the animals display group-level cohesion \cite{puckett2014}. Recently this system has also generated interest for possible indications of critical behavior \cite{giardina2014prl,chate2014insect}. Previous descriptions of insect swarms have accounted for the tight binding of individuals to the group either by introducing a confining potential \cite{giardina2014plos,giardina2014prl} or by invoking external environmental cues \cite{downes1969}. Here, we instead develop a swarm model inspired by the dominant sensory mechanism of the insects, and show that group cohesion can emerge naturally instead of being externally imposed. Swarming species of Chironomid midges, such as those we consider here, are known to be very sensitive to acoustic signals \cite{federova2009}, and are thought to be attracted to swarms by the sound they produce.

We thus make the ansatz that midges accelerate toward the sound produced by others; in essence, we hypothesize that in addition to their known pairwise function, acoustic interactions are the basis for coordinating the large-scale collective behavior of the swarm. The exact structure of the acoustic field produced by a freely flying midge is not known, and to our knowledge there are no direct measurements of its detailed spatial structure. However, in other flying insects (flies, for example) the acoustic field produced has been found to have both monopole and dipole components \cite{sueur2005sound}. Since the monopole field decays more slowly compared to the dipole (and any higher multipole) component, in the model we present we have concentrated on its contribution. The monopole sound intensity falls off according to an inverse-square law, and so this hypothesis results in a effective gravity-like force that promotes group cohesion while still allowing for complex individual motion.

We additionally account for the possibility that the midges' sensory perception may adapt to the overall sound level they experience. Our model is thus not derived from an underlying kinetic theory, as is commonly done in models of collective behavior \cite{vicsek1995}, but rather is explicitly long-range and (because of the adaptivity) many-body; and, as the model bears some similarity to self-gravitating systems, we can exploit our intuition for gravity in analyzing our model.

To benchmark this model, we compare some of its predictions with laboratory measurements of several hundred swarms of the non-biting midge \textit{Chironomus riparius}, and find surprisingly good agreement. Since the acoustic interactions in our model are long-range and adaptivity renders them inherently many-body, our results suggest that the midges process more than just local information, as has also recently been proposed for bird flocks \cite{pearce2014}. We use in this work concepts and techniques from classical $N$-body self-gravitating systems, to explain the collective swarming behavior of insects (midges). Although models of collective behavior abound in the literature, this particular approach has never been applied to the swarming problem. What we find especially appealing about it is both that a gravity-like model can produce very complex behavior from simple interactions, as is expected to occur in collective behavior in biology. At the same time, we introduce a concept to the gravitational physics community that is taken from biology, namely the adaptivity of the sensory mechanism, ending up with a new form of gravitational interaction that we term ``adaptive gravity''.

\section{Experimental Setup}

Before discussing our model and its comparison with our empirical data, we describe our laboratory experiments with midge swarms. The details of these measurements have been discussed elsewhere \cite{kelley2013,puckett2014,puckett2014determining}, and so we only give a brief overview here.

Our self-sustaining colony of \textit{C.~riparius} midges was originally established from egg sacs purchased from Environmental Consulting and Testing, Inc. The colony is maintained in a transparent cubical enclosure, 91 cm on a side, at a constant 22$^\circ$C. The midges are exposed to overhead light on a circadian cycle, with 16 hours of light and 8 hours of darkness per day. When the overhead light turns on and off (corresponding to ``€œdawn''€ and ``€œdusk''), adult males spontaneously form swarms. To promote swarm nucleation and to position the swarm in the enclosure, we use a black felt ``€œswarm marker''€ measuring $30\times30$~cm$^2$ placed just above the development tanks. Further details of our husbandry procedures are given in refs.~\cite{kelley2013} and \cite{puckett2014}.

To quantify the motion of the midges, we record movies of swarming events with three hardware-synchronized Point Grey Flea3 cameras at a rate of 100 frames per second. We have shown previously that this data rate is sufficient to resolve even the acceleration of the midges \cite{kelley2013}. The cameras are arranged in a horizontal plane about 1~m from the center of the swarm with angular separations of approximately 30$^\circ$ and 70$^\circ$. Prior to recording data, the camera system is calibrated using Tsai's model \cite{tsai1987}; subsequently, the two-dimensional coordinates of each midge on each camera (found by simple image segmentation and intensity-weighted averaging) can be combined to find the midge positions in three-dimensional space. The sequences of time-resolved positions are then linked into trajectories using a fully automated multi-frame predictive tracking algorithm \cite{ouellette2006}. For various reasons, trajectories may sometimes be broken; thus, in a post-processing step we link trajectory fragments using Xu's method of re-tracking in a six-dimensional position-velocity space \cite{xu2008}. After trajectory construction, we compute accurate time derivatives by convolving the tracks with a smoothing and differentiating kernel \cite{kelley2013}.

For the results shown here, we analyzed data from 128 swarming events. Although the number of individuals was not uniform from swarm to swarm, we have shown previously that the swarms reach a statistical ``asymptotic'' regime at surprisingly small population sizes \cite{puckett2014determining}; here, the mean number of individuals per swarm was about 10. Finally, for reference below, we note that the body size of a male \textit{C.~riparius} midge is about 7~mm in length. Typical flight speeds of the midges are roughly 0.5~m/s, and peak instantaneous accelerations are on the order of 5~m/s$^2$. The sound produced by a male's beating wings is broadband, but has a fundamental frequency of about 575~Hz, as measured in our experiments \cite{ni2015}. It is difficult to measure the sound amplitude produced by a freely flying midge precisely; within a few body lengths of the midge, we have measured it to be roughly 55~dB.

In previous work, we have analyzed data from these swarms to characterize their dynamics \cite{kelley2013,puckett2014,puckett2014determining,puckett2015}. Without going into detail here, we briefly summarize our primary findings. Although our midge swarms remain confined to a compact region of space with a statistically sharp boundary (in a way that appears to be self-organized \cite{puckett2014determining}), teasing out pairwise interactions within the swarm is very challenging \cite{puckett2014,puckett2015} and the swarms show no net internal order \cite{kelley2013}. At a mean-field level, the statistics of the swarms share some features of an ideal gas in a harmonic trap \cite{kelley2013,puckett2014}. Thus, with our model, we attempt to capture these primary effects: strong binding of individuals to the swarm as a whole but no strong signature of pairwise interaction at the mean-field level, overall disorder inside the swarm, and complex individual trajectories.

\section{Model}

In the absence of viscous damping, the intensity of the monopole component of the sound produced by an isotropic point emitter decays purely geometrically: since the acoustic energy in the spherical wavefront is fixed and the surface area of these wavefronts grows as $r^2$, where $r$ is the distance from the emitter, the sound intensity must fall off as $1/r^2$. The typical fundamental frequency produced by the wingbeats of a male midge in our experiments is roughly 575~Hz \cite{ni2015}, corresponding to a wavelength of about 60~cm. Given the typical size of a swarm (diameters of roughly 20~cm in our experiments) and nearest-neighbor distance (about $4-7$~cm \cite{kelley2013}), midges in a swarm are therefore in the near-field regime of the acoustic field of their neighbors. In refs.~\cite{lighthill1962sound,bennet1998size}, the decay of the pressure field due to the wing-beat of a fly was measured, and it was found that both in the near-field and far-field the decay follows a $1/r$ behavior. This results in a decay of the intensity as $1/r^2$, which is the form we use in our model. We note that due to interference effects between the incoherent sources, the acoustic field decays roughly as a Gaussian for length scales larger than the wavelength, i.e. $\gtrsim60$~cm; this size is, however, much larger than the typical swarms we have studied.

Next, we make the hypothesis that an individual midge accelerates towards a neighbor via an effective ``force'' that is proportional to the sound intensity. Given the estimates above, this means that the force between a pair of midges $i$ and $j$ separated by a distance $r_{ij}=| \vec{r}_i-\vec{r}_j |$ will scale as $1/r_{ij}^2$, just as the gravitational attraction between a pair of point masses would. At present, we must treat this choice purely as an ansatz, as the details of the form of any pairwise interactions between midges is very difficult to access experimentally \cite{puckett2014,puckett2015}. However, the assumption the midge response to acoustic signals is an acceleration towards the sound source is the simplest choice one can make. Choosing the response to be at the velocity level would be somewhat unnatural, since velocity cannot be directly controlled by the insects: changes to the velocity must come from forces applied by the insect, and therefore accelerations. And strong (albeit indirect) experimental support for this assumption comes from the observation of a net linear restoring force acting towards the swarm center (Fig.~\ref{forceShape}b). The \emph{only} form of binary interactions that gives this linear restoring force towards the swarm center is an inverse-square force relation.

For many animals, the perception of sound is not fixed, but rather adapts to the total sound intensity so that acoustic sensitivity drops when there is strong background noise. This is a common feature of biological sensory organs, preventing their damage and saturation. We thus make a second ansatz: that in general the midges' acoustic perception adapts to the overall sound level, and that specifically it follows the fold-change detection mechanism \cite{shoval2010}, which is ubiquitous in nature. In that case, the effective force on midge $i$ due to midge $j$ is given by
\begin{equation}
\vec{F}^{i}_{eff}=C\sum_{j}\hat{r}_{ij}\frac{1}{|\vec{r}_i-\vec{r}_j|^2}\left(\frac{R_{ad}^{-2}}{R_{ad}^{-2}+\sum_{k}|\vec{r}_i-\vec{r}_k|^{-2}}\right) \label{feffmany},
\end{equation}
where $\hat{r}_{ij}$ is the unit vector pointing from midge $i$ to midge $j$, $C$ is a constant with dimensions of $mass \cdot length^3/time^{2}$, and $R_{ad}$ is the length scale over which adaptivity occurs. In other words, when a single midge is closer than $R_{ad}$ the sound it emits is strong enough that the receiving midge needs to adapt its sensitivity to reduce the perceived signal. Beyond this distance there is no need for such adaptivity for the sound of a single midge.

Equation (\ref{feffmany}) constitutes the core of our ``adaptive gravity'' model (AGM) for the acoustic interactions of the midges. We note that this model assumes that the midges can sense both the intensity and the direction of the sound produced by others; it is thought, however, that the specialized Johnston's organs of male swarming insects are indeed able to do so \cite{cator2010}. We also note that we are making the simplifying assumption that each midge is identical; although this assumption is certainly not fully accurate, it should allow us to make reasonable mean-field predictions.

With the assumption of adaptivity, the force felt by each midge is inherently many-body and cannot be written as a sum of two-body interactions (Eqs.~(\ref{normalized force}) and (\ref{feffmanysum})) due to the sum over all the midges that appears in the denominator of the adaptivity factor (Eq.~(\ref{feffmany})). Thus, in this formulation, every midge feels a force that contains global, long-range information about the swarm, but this force cannot be parsed to distinguish the effects of any single neighbor. Thus, in this AGM the force that binds individual midges to the swarm is truly an emergent, group-level property that arises naturally from within the swarm without any appeal to external effects.

To build intuition for the behavior of this model, let us consider two limits. For $r_{ij}\gg \sqrt{N}R_{ad}$ (that is, when the distance between a pair of midges far exceeds the range of adaptivity, and $N$ is the number of midges in the swarm), the effective force reduces to a purely gravitational interaction and becomes
\be \label{gravitational}
\vec{F}^{i}_{eff,g} \rightarrow C\sum_{j}\hat{r}_{ij}\frac{1}{|\vec{r}_i-\vec{r}_j|^2}.
\ee
In the opposite limit, when $r_{ij} < \sqrt{N}R_{ad}$, the adaptive nature of the acoustic sensitivity becomes dominant. In that case, the adaptivity simply reduces to a rescaling of the sound perceived by each midge by the total buzzing noise amplitude, and the effective force becomes
\be
\vec{F}^{i}_{eff,a} \rightarrow \frac{C}{R_{ad}^2}\cdot\frac{\sum_{j}\hat{r}_{ij}\frac{1}{|\vec{r}_i-\vec{r}_j|^2}}{N_{tot}}, \label{normalized force}
\ee
where the total buzzing noise amplitude at $\vec{r}_i$ is $N_{tot}=\sum_{j}(|\vec{r}_i-\vec{r}_j|)^{-2}$.

In pure gravity, the potential is additive, and the principle of superposition applies. Due to adaptivity, however, this property is lost in our model. That is, the effective potential felt by a midge due to many other midges is \emph{not} the sum of two-body interactions using Eq.~(\ref{ueff}). This can be seen by considering many interacting midges. The effective force felt by midge $i$ due to the others (indices $j$) is given in Eq.~(\ref{feffmany}) which is \emph{not} equal to the sum over two-body forces (see Eq.~(\ref{feff}) in the Appendix), which would be
\begin{equation}
\vec{F}^{i}_{eff,2}=C\sum_{j}\hat{r}_{ij}\frac{1}{|\vec{r}_i-\vec{r}_j|^2}\frac{R_{ad}^{-2}}{R_{ad}^{-2}+|\vec{r}_i-\vec{r}_j|^{-2}}. \label{feffmanysum}
\end{equation}

Unlike in pure gravity, the forces on midge $i$ due to others are not additive and the superposition principle does not apply, since the total buzzing noise term does not depend on direction. As a consequence, there are no conservation laws in such a system (except mass conservation), and the sum of forces felt by all the midges within an isolated swarm need not vanish, as it must in regular gravity. Thus, in this model the center of mass of the swarm can experience accelerations; so, even though the AGM naturally leads to swarm cohesion, one may need to posit external effects to prevent drift of the swarm as a whole.

\section{Results}
\subsection*{Effective spring constants: Spherical swarms.}
Gauss's law for gravity states that the gravitational flux through a closed surface is proportional to the enclosed mass \cite{greiner2004classical}. In our analogy, each midge has an effective unit ``mass,'' and therefore Gauss's law for the force in the ``pure gravity'' regime (Eq.~(\ref{gravitational})) is
\be
\int_{\partial V}\!\!\vec{F}_{eff}\cdot d\vec{A}=-4\,\pi\,C\,\int_{V}\!\!\rho(\vec{r})d^3r, \label{gauss}
\ee
where $V$ is a three-dimensional volume, $\partial V$ is its boundary, $d\vec{A}$ is a surface element, and $\rho(\vec{r})$ is the density of midges.

\begin{figure}[tb]
\centering
\includegraphics[width=0.8\linewidth]{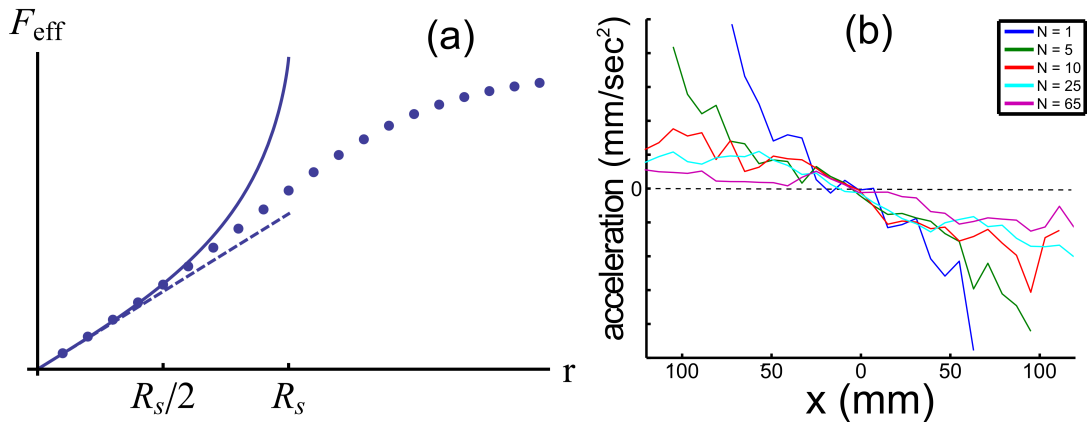}
\caption{\label{forceShape} Mean-field effective forces. a. Calculated mean effective force acting on a midge within a spherical swarm as a function of the radial position $r$. The solid line shows the force due to adaptive gravity (Eq.~(\ref{feffmany})) for a swarm of uniform density and radius $R_s$, compared to the case of pure gravity (dashed line). When the density has a Gaussian profile (with spatial variance $R_s/2$, Eqs.~(\ref{rhoGauss})-(\ref{NtotGauss})), the adaptive-gravity interactions give rise to the force shown with the dotted line. b. Measured mean acceleration as a function of position for laboratory swarms with different numbers of midges (shown with different colors), showing the roughly linear behavior near the swarm center.}
\end{figure}

We begin with a spherical swarm of uniform density $\rho$ and radius $R_s=\langle r \rangle$. The characteristic value of the total buzzing noise intensity at the origin is $N_{tot}\sim N\,/\, R_{S}^{2}$, where $N$ is the number of midges. Thus, when $R_S\gg \sqrt{N} R_{ad}$ we are in the pure gravity regime. In this case, from the analog of Gauss's law (Eq.~(\ref{gauss})), the force is restoring and linear with respect to the distance $\vec{r}$ of a midge from the center of the swarm (Fig.~\ref{forceShape}a), and is given by
\be \label{linear force}
\vec{F}_{eff}=-\frac{4\,\pi\,C\,\rho}{3}\vec{r}.
\ee
Since this force is harmonic (that is, restoring and linear in $\vec{r}$), we can characterize its strength with an effective ``spring constant'' $K=4\,\pi\,C\,\rho/3$. We stress that this behavior is unique for $\hat{r}_{ij}/r_{ij}^2$ interactions, assuming that the motion arises only from interactions between the midges. Previously, we found that the average acceleration of midges in laboratory swarms also has this harmonic form \cite{kelley2013}, providing support for our model. However, in these laboratory swarms, the spring constants were found to depend on the swarm size $R_s$ (Fig.~\ref{forceShape}(b)), unlike in pure gravity.

For swarms with large numbers of individuals, however, our model enters its adaptive regime, where $R_S < \sqrt{N} R_{ad}$. In this regime, the net force is still linear and restoring; but due to the adaptive terms in Eq.~(\ref{feffmany}), the spring constant $K$ will depend on the swarm size $R_s$. To leading order in $r/R_{s}$, the model predicts that $K \propto (R_{ad}^2\,R_{s})^{-1}$.
This result is derived in the Appendix (Eqs.~(\ref{New_grav})-(\ref{Ntot})), and also follows from Eq.~(\ref{normalized force}) using simple dimensional analysis, since
\bea
\left(\sum_{j}\frac{R_{ad}^2}{r_{ij}^2}\right)^{-1}\rightarrow
\left(R_{ad}^2\,\int_{0}^{R_s}\frac{d^{3}r}{r_{ij}^2}\right)^{-1}
\sim \left(R_{ad}^2\,R_{s}\right)^{-1}. \label{k dependence}
\eea

\begin{figure}[tb]
\centering
\includegraphics[width=\linewidth]{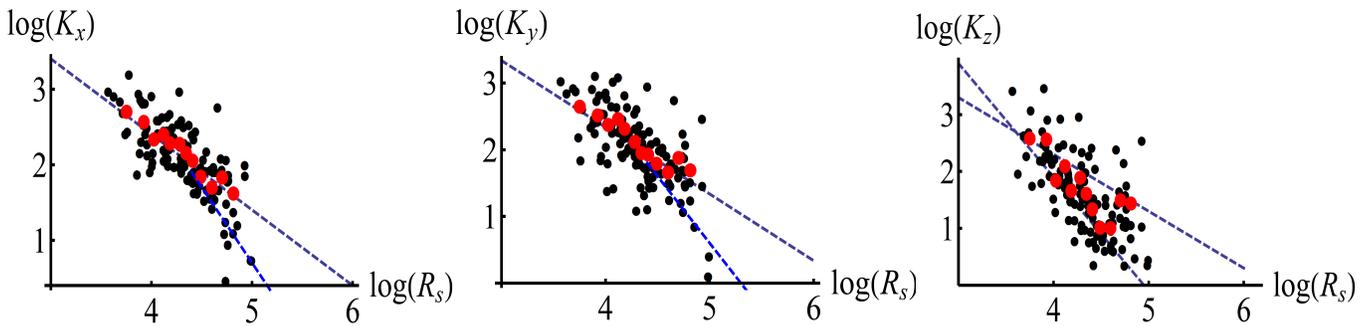}
\caption{\label{KvsRs} Effective spring constants in the horizontal directions ($K_x$ and $K_y$) and the vertical direction ($K_z$) as a function of the average swarm radius $R_s$, plotted on logarithmic axes. The black circles denote the raw data for each swarm, the red circles show binned averages, and the dashed lines denote the $R_s^{-1}$ scaling predicted by the AGM (Eq.~(\ref{k dependence})). We also plot the $R_s^{-2}$ scaling predicted for cylindrical swarms (regular gravity, Eq.~(\ref{KzCylinder})), which seems to fit the behavior of larger swarms. Note that in the vertical $z$-direction the data is more scattered, which may be due to Earth's gravity.}
\end{figure}

When we examine the experimental data for $K_x$ and $K_y$ (where $x$ and $y$ are in the plane and $z$ is vertical), we find good agreement with the model prediction that $K \propto R_s^{-1}$ (Fig.~\ref{KvsRs}a,b). This agreement is a consequence of the roughly constant density in swarms of different sizes (except for small swarms of fewer than $\sim10$ midges \cite{puckett2014determining}), and gives a lower bound on $R_{ad}\gtrsim45mm$ since the adaptive regime applies to the smallest swarms. For $K_z$ we find a decrease that is faster than predicted (Fig.~\ref{KvsRs}c), as discussed further below.

We note that away from the swarm center the adaptive-gravity interaction gives rise to a restoring force that deviates from the form of pure gravity even for a uniform density swarm (Fig.~\ref{forceShape}a). Thus, we also calculated the force for a swarm with a Gaussian density profile, as was observed in experiments \cite{kelley2013}(Fig.~\ref{forceShape}a), and find that it is roughly linear over the entire swarm size, but saturates at large radii.

The calculation of the adaptive force near the swarm center, for a Gaussian density profile, is as follows. We take a Gaussian density profile with width $\sigma_s$,
so that the density in cylindrical coordinates is given by
\be\label{rhoGauss}
\rho(r,z)=\frac{\exp{(-\frac{r^2+z^2}{2\,\sigma_s^2}})}{(2\,\pi)^{\frac{3}{2}}\,\sigma_S^3}.
\ee
The gravitational force at $z_0$ is then
 \be
F_{eff,g}(z_0)=2\,\pi\,\int_{-\infty}^{\infty}\!\!\!dz'\int_{0}^{\infty}\!\!\!r'dr'\rho(r',z')\,\frac{z'-z_0}{\left[r'^2+(z'-z_0)^2\right]^{\frac{3}{2}}}
,
\ee
and the total buzzing noise is
\be\label{NtotGauss}
N_{tot}(z_0)=2\,\pi\,\int_{-\infty}^{\infty}\!\!\!dz'\int_{0}^{\infty}\!\!\!\rho(r',z')\frac{r'dr'}{r'^2+(z'-z_0)^2}.
\ee
The integrals were solved numerically using $Mathematica\,\,\,
9.0$, and were used to plot the dotted line in Fig.~\ref{forceShape}a.

To conclude this part, we find that the accelerations of midges near the swarm center follow the linear relation expected from gravity-like interactions. Furthermore, the effective spring constant decreases with swarm size, exactly as predicted by adaptivity. The effects of non-spherical swarms are dealt with next.

\subsection{Effective spring constants: Ellipsoidal swarms.}

\begin{figure}[tb]
\centering
\includegraphics[width=0.8\linewidth]{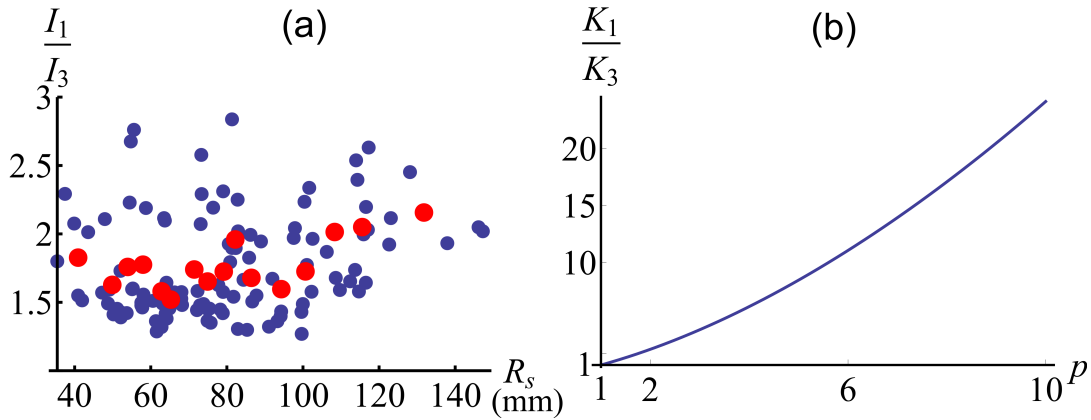}
\caption{(a) Observed average inertia eigenvalue ratio $I_1/I_3$ as a function of the swarm size $R_s$ (blue circles: raw data for each swarm, red circles: binned average). For smaller swarms, the ratio is $\sim1$ which is the spherical limit, while for larger swarms this ratio is larger than 1 due to the elongation of the swarm in the $z$-direction ($I_3>I_1$, Fig.~\ref{fig:v1}a). (b) Calculated spring constant ratio $K_1/K_3$ for a prolate, axisymmetric swarm, using Eq.~(\ref{kratioProlate}), as a function of the aspect ratio $p\equiv a/c$ (solid line). For small $p$ the behavior is linear, as given in Eq.~(\ref{kratioProlateLin}).}
\label{prolateSwarm}
\end{figure}

In the measured swarms, the effective spring constant in the vertical ($z$) direction is consistently smaller than those in the horizontal ($x,y$) directions \cite{kelley2013}; additionally, it is also observed to decrease faster with swarm size than is predicted by our AGM for spherical swarms (Fig.~\ref{KvsRs}c). For real swarms, the $z$ direction differs from the $x$ and $y$ directions in several ways. First, along this direction midges are affected by the earth's gravitational pull. Additionally, swarms tend to form over visual features on the ground \cite{puckett2014determining}, which breaks the isotropic symmetry. Empirically, all these differences tend to cause larger swarms to elongate along the $z$ axis \cite{kelley2013,puckett2014determining} (Fig.~\ref{prolateSwarm}a). As we calculate below (and show in Fig.~\ref{prolateSwarm}b), for swarms that are elongated along the $z$-axis, our model predicts that the effective spring constant in the $x,y$-plane ($K_{1,2}$) is larger than in the $z$-direction ($K_3$). We therefore attribute the observed smaller spring constant in the $z$-direction for larger swarms (Fig.~\ref{KvsRs}) to the elongation of the swarms along the vertical axis. Furthermore, we can calculate the scaling of $K_3$ with swarm size in the limit of a highly elongated (cylindrical) swarm, such that it has a fixed radius $R$ in the $xy$-plane and a variable length $L\gg R$ along the $z$-axis. In this limit we find the scaling $K_3\propto R_s^{-2}$ (Eq.~(\ref{KzCylinder})), as denoted in Fig.~\ref{KvsRs}c. Note that in this limit, we are beyond the perfect adaptivity regime, and the scaling result for $K_3$ is identical to that of pure gravity.

\begin{figure}[tb]
\centering
\includegraphics[width=0.8\linewidth]{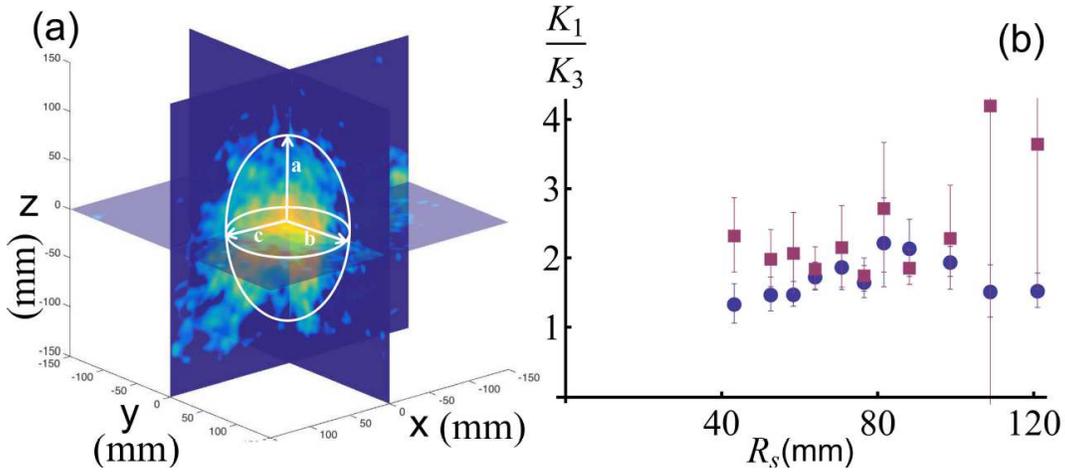}
\caption{\label{fig:v1} Swarm shape. a. Example of the shape of a laboratory swarm of $N=7$ midges, which is elongated along the vertical ($z$) direction. Also shown is the ellipsoid that approximates the swarm shape, which by construction has the same ratios of inertia eigenvalues. b. The ratio of the effective spring constants $K_{x,y}/K_{z}$ as a function of the swarm radius $R_s$, using the data in Table 1. Measured values are shown in blue, and those calculated from the model are shown in purple.}
\end{figure}

In Fig.~\ref{fig:v1}a we show the shape of a typical elongated swarm. We treat the swarm shape using an ellipsoidal approximation to refine the analysis of the effective spring constants along the different directions. We assume that the swarm is an ellipsoid with semi-axes $a$, $b$, and $c$, where $c<b<a$ (see Fig.~\ref{fig:v1}a), along the $x,y,z$-axes. The effective spring constants are then given by (see Appendix, Eqs.~(\ref{MainEllipsoid})-(\ref{NtotEllipse}))
\bea
K_1=\pi\,a\,b\,c\,\rho\int_0^{\infty}\frac{dv}{(c^2+v)\sqrt{\beta(v)}}, \label{spring1}
\\
K_2=\pi\,a\,b\,c\,\rho\int_0^{\infty}\frac{dv}{(b^2+v)\sqrt{\beta(v)}}, \label{spring2}
\\
K_3=\pi\,a\,b\,c\,\rho\int_0^{\infty}\frac{dv}{(a^2+v)\sqrt{\beta(v)}}, \label{spring3}
\eea
where $\beta(v)\equiv (a^2+v)(b^2+v)(c^2+v)$ and $K_1>K_2>K_3$ since $c<b<a$. Eqs.~(\ref{spring1})--(\ref{spring3}) relate the effective forces in the swarm to its overall shape. The measured values of the spring constants and shapes of the swarms are summarized in Table 1. We characterize the shapes by the ratios of the moments of inertia-tensor eigenvalues $\eta_1=I_1/I_2$ and $\eta_2=I_1/I_3$. As part of our ellipsoid approximation, we assume that the inertia eigenvectors are oriented along the principal axes of the ellipsoid and that each inertia tensor eigenvalue corresponds to one of the axes.  Let us assume that $I_1>I_2>I_3$, without loss of generality. Then
\bea
I_1=\frac{4\,\pi}{15}\rho\,a\,b\,c \(a^2+b^2\),\label{I1}\\
I_2=\frac{4\,\pi}{15}\rho\,a\,b\,c \(a^2+c^2\),\label{I2}\\
I_3=\frac{4\,\pi}{15}\rho\,a\,b\,c \(b^2+c^2\)\label{I3},
\eea
and
\bea
\eta_1\equiv\frac{I_1}{I_2}=\frac{a^2+b^2}{a^2+c^2}, \label{InertiaRat1}\\
\eta_2\equiv\frac{I_1}{I_3}=\frac{a^2+b^2}{b^2+c^2}. \label{InertiaRat2}
\eea
From Eqs.~(\ref{InertiaRat1}, \ref{InertiaRat2}), one can express the parameters of the ellipsoid as a function of $\eta_1$ and $\eta_2$:
\bea
\frac{b^2}{c^2}&=&\frac{\eta_1+(\eta_1-1)\,\eta_2}{\eta_1+\eta_2-\eta_1\,\eta_2},\label{bc}
\\
\frac{a^2}{c^2}&=&\frac{\eta_1-(\eta_1+1)\,\eta_2}{\eta_1\(\eta_2-1\)-\eta_2},\label{ac}
\eea
and then the ratios $K_1/K_2$ and $K_1/K_3$ can be obtained from Eqs.~(\ref{spring1})-(\ref{spring3}).
Note that in this analysis (Table 1) the direction of each $K_i$ can be different, as they are defined by their relative strength so that $K_1>K_2>K_3$. In all cases, however, the smallest effective spring constant is in the $z$ direction ($K_3=K_z$), since we always observe that swarms are stretched in the vertical direction (Fig.~\ref{prolateSwarm}).

\begin{table}[h]
\caption{Experimental data for the dependence of the mean swarm shape (given by the inertia eigenvalues $I_1,I_2,I_3$) and effective spring constants along the principle directions, on the swarm size given by the average radius $R_s$ (binned averages).}
\begin{tabular}{|c|c|c|c|c|c|}
  \hline
  $R_s$(mm) & $K_1$(1/sec$^2$) & $K_2$(1/sec$^2$) & $K_3$(1/sec$^2$)

   & $\eta_1=I1/I2$ & $\eta_2=I1/I3$ \\
  \hline
  \label{Table1}
43.22 &  15.83 & 14.72 & 11.80  & 1.38 & 1.81 \\
52.58 & 12.33 & 11.06 & 8.34 & 1.27 &1.70 \\
58.40 & 10.62 & 9.50 & 7.17 & 1.37 & 1.68 \\
63.97 & 10.12 & 10.06 & 5.83 & 1.26 & 1.61 \\
70.77 & 9.09 & 7.98 & 4.84 & 1.32 & 1.77 \\
76.53 & 8.83 & 7.60 & 5.33 & 1.25 & 1.54 \\
81.59 & 7.36 & 6,55 & 3.31 & 1.40 & 1.98 \\
88.15 & 6.65 & 6.11 & 3.10 & 1.26 & 1.62 \\
98.62 & 5.94 & 5.49 & 3.05 & 1.43 & 1.74 \\
108.89 & 6.94 & 6.34 & 4.56 & 1.64 & 2.07 \\
120.99 & 4.82 & 4.62 & 3.15 & 1.57 & 2.06 \\
\hline
\end{tabular}
\end{table}

The ellipsoid parameters can be expressed in terms of the ratios of the inertia-tensor eigenvalues (Eqs.~(\ref{I1})-(\ref{ac})), thereby relating the effective spring constant ratios $K_1/K_2$ and $K_1/K_3$ to the measured shape. We plot these ratios for both measured swarms and for the model in Fig.~\ref{fig:v1}b. For smaller swarms, we find good agreement between the theoretical and measured values, where the discrepancies are primarily due to misalignment of the inertia-tensor eigenvectors with the principal axes of the ellipsoid. For the largest swarms, however, there is a significant deviation between the two. In those cases, the ellipsoidal approximation may not be valid, as we sometimes observed a tendency for these large swarms to split into a main body and satellite swarm and thus violate the assumptions of the model.

In the case of a prolate axisymmetric ellipsoid $b=c<a$, we have
$K_2=K_3$ and from Eqs.~(\ref{spring1})-(\ref{spring3}) we get
\be
\frac{K_1}{K_3}=\left.\int_0^{\infty}\!\!\!\frac{dv}{(1+v)^2\,\(p^2+v\)^{\frac{1}{2}}}\middle/\int_0^{\infty}\!\!\!\frac{dv}{(1+v)\,\(p^2+v\)^\frac{3}{2}}\right.=\frac{p\(p-p^3+\sqrt{p^2-1}\,\cosh^{-1}
p\)}{2\,\(p^2-1-p\sqrt{p^2-1}\cosh^{-1}{p}\)},
\label{kratioProlate}
\ee
where $p\equiv a/c$. For small deviations from spherical symmetry
$p=1+\epsilon$,~$\epsilon \ll 1$ we have
\be
\frac{K_1}{K_3}=1+\frac{6}{5}\epsilon+\mathcal{O}(\epsilon^2).
\label{kratioProlateLin}
\ee
This result is shown in Fig.~\ref{prolateSwarm}b.

To summarize the results of this part, our AGM explains why the elongation of the swarms along the vertical axis gives rise to lower effective spring constant that is observed in this direction, as well as to the different scaling with the swarm size (Fig.~\ref{KvsRs}).

\subsection{The virial relation.}

Despite the fact that adaptivity prevents us from formulating many conservation laws, we can still develop an analog to the virial theorem, based on mass conservation. We can define analogous kinetic and potential energies, and write a conservation law that relates them in case of stationarity. For this purpose we follow here Chandrasekhar \cite{Chandrasekhar} and ideas that are used in galactic dynamics (see for instance \cite{binney2011galactic}) to derive the virial equations for an ideal self-gravitating fluid described in terms of density $\rho(\vec{r},t)$ and an isotropic pressure $p(\vec{r},t)$. The virial equations that we derive here are equations of the second order, since they relate second-order moments. There are equations of higher order as well. Later we will show how this derivation is related to the midge swarms.

We assume that, in addition to pressure gradients, the only fields that act on the fluid are the internal self-gravitational field of the fluid $\vec{g}^{(in)}(\vec{r})$ and the external gravity field in the $z$ direction, given by
\be
\vec{g}^{(ext)}=-g\hat{z}.
\ee
Then the hydrodynamic equation governing the motion of the fluid is given by
\be \label{hydro}
\rho \frac{dv_{i}}{dt}=-\frac{\partial p}{\partial r_{i}}+\rho\,g_{i}(\vec{r}),
\ee
where $v_i(\vec{r},t)$ is the velocity of the fluid,
\[\frac{d}{dt}=\frac{\partial}{\partial t}+v_{j}\frac{\partial}{\partial r_{j}}\]
is the total time derivative (typically called the material derivative in fluid mechanics) and
\be \label{full field}
g_{i}(\vec{r})=g^{(in)}_{i}(\vec{r})-g\delta_{iz}.
\ee
In order to obtain the second order virial equations we multiply Eq. (\ref{hydro}) by $r_j$ and integrate over the entire volume $V$.  Integration of the term on the left hand side of Eq. (\ref{hydro}) gives
\be \label{firstInt}
\int_{V}\rho\,r_j\,\frac{dv_{i}}{dt}d^3\vec{r}=\int_{V}\rho\left[\frac{d}{dt}(r_j\,v_i)-v_j\,v_i\right]\,d^3\vec{r}=\frac{d}{dt}\(\int_{V}\rho\,v_i\,r_j\,d^3\vec{r}\)-2\,T_{ij},
\ee
where
\be
T_{ij}=\int_{V}\rho\,v_{i}v_{j}\,d^{3}\vec{r},
\ee
is the kinetic energy tensor. The last equality of Eq. (\ref{firstInt}) is a direct consequence of mass conservation, namely
\be
\frac{d}{dt}\(\int_{V}\rho(\vec{r},t ) d^{3}\vec{r}\)=0.
\ee

The same procedure with the first term on the right hand side of Eq. (\ref{hydro}) gives
\be
\int_{V}r_j\frac{\partial p}{\partial r_i}d^{3}\vec{r}=\delta_{ij}\int_{V}p\,d^{3}\vec{r}\equiv\delta_{ij}\Pi,
\ee
where we use integration by parts (assuming that the fluid is concentrated in a bounded region of space), and $\Pi$
is the total internal pressure of the fluid.
The last term of Eq. (\ref{hydro}) gives us a ``potential energy'' term:
\be
W_{ij}=\int_{V}\rho\,r_{j}g_{i}(\vec{r})d^{3}\vec{r}.
\ee
Collecting the terms together we get
\be \label{firstDerEq}
\frac{d}{dt}\(\int_{V}\rho\,v_i\,r_j\,d^3\vec{r}\)=2\,T_{ij}+W_{ij}+\delta_{ij}\Pi.
\ee
We can write Eq. (\ref{firstDerEq}) in a form similar to Newton's second law by introducing the moment of inertia tensor
\be
I_{ij}=\int_{V}\rho\,r_i\,r_{j}d^{3}\vec{r},
\ee
and since the tensors on the right hand side of Eq. (\ref{firstDerEq}) are symmetric we get
\be
\frac{1}{2}\frac{d^2\,I_{ij}}{dt^2}=2\,T_{ij}+W_{ij}+\delta_{ij}\Pi.
\ee
When the system is stationary the moment of inertia does not change over time, and we arrive at the tensor form of the virial theorem:
\be
2\,T_{ij}+W_{ij}+\delta_{ij}\Pi=0. \label{virial}
\ee

\begin{figure}[h]
     \includegraphics[width=1.0\linewidth]{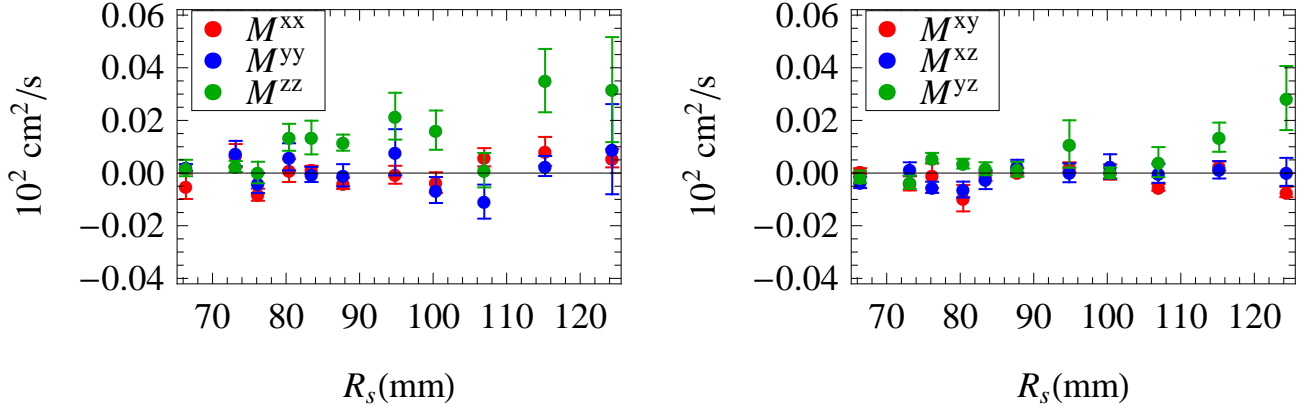}
\caption{The six components of the tensor $\bar{M}^{ij}$ (Eq. (\ref{Mtensor})) that captures deviations from stationarity, as a function of the swarm size $R_s$}
\label{stationarity.fig}
\end{figure}

In order to apply the tensor virial equations presented above to the midge swarms, we will have to write its discrete analogues for $N$ particles with equal (unit) masses. The moment of inertia tensor is now
 \be
\bar{I}^{ij}\equiv\frac{1}{N}\sum_{n=1}^{N}r_n^{i}r_{n}^{j},
\ee
where the bar denotes an average value per midge.
Its derivative with respect to time can give us an indication for deviations of the system from stationarity, namely
\be
\bar{M}^{ij}\equiv\frac{d\bar{I}^{ij}}{dt}=\frac{1}{2\,N}\sum_{n=1}^{N}\(r_n^{i}v_{n}^{j}+v_{n}^{i}r_{n}^{j}\).
\label{Mtensor}
\ee
We use upper indices for the quantities that are defined with discrete summation. In Fig.~\ref{stationarity.fig} we show the values of $\bar{M}^{ij}$ taken for swarms of different sizes. Out of 126 measured swarms, we consider in this section binned data from 69 swarms that consisted of five midges or more, since for swarms with too few midges the average is meaningless. In addition we take time averages of the quantities over roughly one minute, so that the swarm is approximately in a steady state. The average values of the different components of $\bar{M}^{ij}$ are small compared to the typical angular momentum, which is two orders of magnitude larger. We therefore conclude that the midge swarms are stationary, and we therefore expect that the virial relation (Eq. (\ref{virial})) should hold. Deviations from stationarity might occur due to influx or outflux of midges (negligible) or irreversible processes. The small increase in the values of $\bar{M}^{zz}$ and $\bar{M}^{yz}$ for large swarms might be an indication for such an irreversible process, such as fragmentation as a result of the elongation along the vertical direction.

\begin{figure}[h]
     \includegraphics[width=0.6\linewidth]{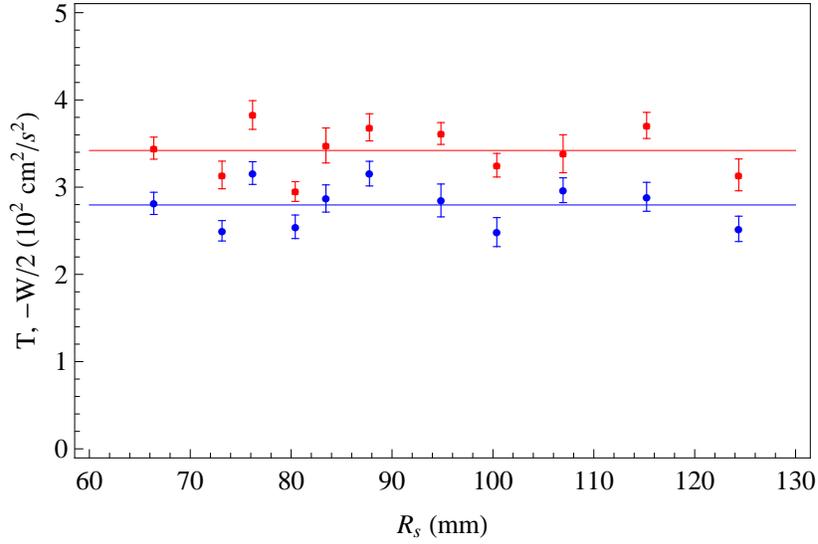}
\caption{The mean kinetic energy $\bar{T}$ (red) and half of the mean potential energy $-\bar{W}/2$ (blue) as a function of the swarm size $R_s$ (Eq.~(\ref{scal virial})).}
\label{Energiesfar.fig}
\end{figure}

Given the stationarity that we have found in the swarms, we now test the validity of the virial relation (Eq. (\ref{virial})), using its discrete analogue. The mean kinetic energy tensor of a midge in the swarm is
\be
\bar{T}^{ij}\equiv\frac{1}{2\,N}\sum_{n=1}^{N}v_n^{i}v_{n}^{j},
\ee
and
\be
\bar{W}^{ij}\equiv\frac{1}{2\,N}\sum_{n=1}^{N}r_n^{i}F_{n}^{j}+F_{n}^{i}r_{n}^{j}
\ee
is its ``potential energy'' tensor. The discrete virial equation is therefore
\be
2\,\bar{T}^{ij}+\bar{W}^{ij}+\delta^{ij}\bar{\Pi}=0, \label{dis virial}
\ee
where $\bar{\Pi}$ is the average pressure on a midge. This pressure term represents the effect of the mutual repulsion when the midges arrive too close to each other. It is easier to interpret and check the trace of the tensorial equation (\ref{dis virial}), namely the scalar form of the virial equation
\be
2\,\bar{T}+\bar{W}+3\bar{\Pi}=0. \label{scal virial}
\ee
Here $\bar{T}$ is the mean total kinetic energy of a midge, $\bar{W}$ is its mean ``potential energy'', and $\bar{\Pi}$ is the mean isotropic pressure on a midge.
In Fig.~\ref{Energiesfar.fig} we show the measured $\bar{T}$ and $-\bar{W}/2$ for different swarm sizes, when integrating over all the midges in the swarm. We can see that they are approximately constant as functions of $R_{s}$ and their mean values are
\bea
\langle\bar{T}\rangle=(3.42\pm0.08)\cdot 10^2\quad cm^2/s^2,\nonumber
\\
\langle-\bar{W}/2\rangle=(2.80\pm0.08)\cdot 10^2\quad cm^2/s^2.
\eea
Therefore, according to Eq.~(\ref{scal virial}), the difference between the two gives the mean pressure in the swarm:
\be
\bar{\Pi}=-(0.41\pm 0.07)\cdot 10^2 cm^2/s^2.
\ee

\begin{figure}[h]
     \includegraphics[width=1.0\linewidth]{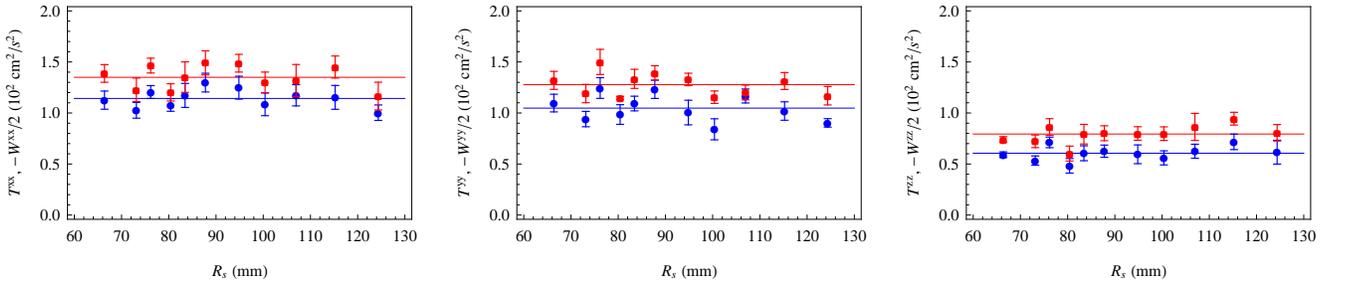}
\caption{The diagonal components of the kinetic $\bar{T}^{ii}$ (red) and potential $-\bar{W}^{ii}/2$ (blue) energy tensors.}
\label{diagonalEnergies.fig}
\end{figure}

In order to confirm that the identification of the mean pressure is correct, we consider each diagonal component of Eq.~(\ref{dis virial}) separately as is shown in Fig.~\ref{diagonalEnergies.fig}. From the mean values we get that
\bea
 2\,\bar{T}^{xx}+\bar{W}^{xx}=(0.42\pm 0.03)\cdot 10^2 cm^2/s^2, \nonumber\\
 2\,\bar{T}^{yy}+\bar{W}^{yy}=(0.46\pm 0.04)\cdot 10^2 cm^2/s^2,
 \nonumber\\
  2\,\bar{T}^{zz}+\bar{W}^{zz}=(0.38\pm 0.03)\cdot 10^2 cm^2/s^2.
\eea
Since these values are roughly equal it gives us a good confirmation for the \emph{isotropic} origin of the pressure term in the virial relation. Note that the observation that $\bar{W}$ is independent of the swarm size when calculated over the observed density profile \cite{kelley2013} of the whole swarm is in agreement with a calculation done using regular gravity \cite{workinprogress}.

In addition we see from Fig.~\ref{diagonalEnergies.fig} that the mean values of kinetic and potential energies, which are related to the movement in the $z$ direction, are significantly lower than the $x$ and $y$ directions. This is a result of the external gravitational force in this direction that enters the equations (see Eq.~(\ref{full field})). From the point of view of the midge, it seems that it is more beneficial to respond to the effective pull of neighboring midges, rather than waste energy moving up and down against gravity (which holds no information regarding location within the swarm).

\begin{figure}[h]
     \includegraphics[width=1.0\linewidth]{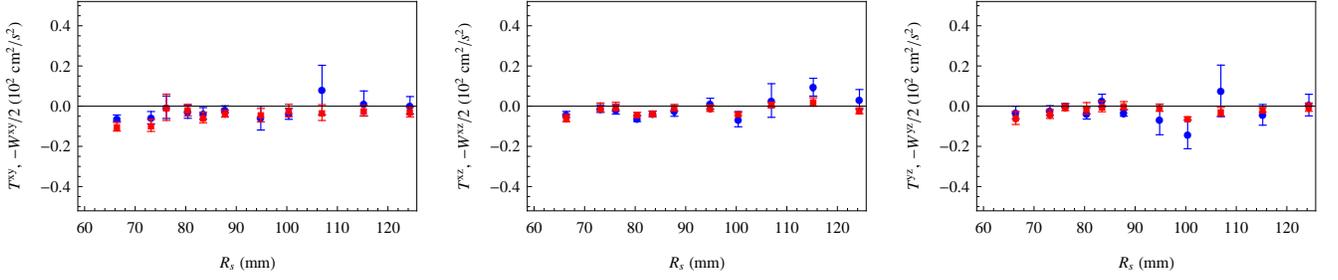}
\caption{The off-diagonal components of kinetic $\bar{T}^{ij}$ (red) and potential $-\bar{W}^{ij}/2$ (blue) energy tensors.}
\label{OffDiagonal.fig}
\end{figure}

The off-diagonal components of the tensors $T^{ij}$ and $W^{ij}$ are roughly null (compared to the diagonal ones) as we show in Fig.~\ref{OffDiagonal.fig}. This is expected for a system without dissipation, which maintains stationarity. Note that on the ``microscopic'' level of each midge, this system is obviously dissipative and out of equilibrium as the midge consumes chemical energy to power its flight. However, we find that on the coarse-grained scale of equivalent particles and forces, the system is effectively dissipation-less.

So far, we have not considered the adaptive nature of the interactions.
Let us assume a uniform density spherical swarm and first calculate the dependence on $R_s$ without adaptivity.
In order to calculate the behavior of the potential energy with $R_s$, we consider again the continuous version of the mean potential energy
\be
\bar{W}=\left.\sum_{k=1}^{3} \int \!\!\! \rho(\vec{r})\,r_k \(F_{ad}\)_{k}d^{3}\vec{r}\right/ \int \!\!\! \rho(\vec{r})d^{3}\vec{r},
\label{Wswarm}
\ee
where $r_k$ is the k-th component of the vector $\vec{r}$.

In the case of a uniform spherical symmetric swarm we have
\be \label{pot_sphe}
\bar{W}=\left. \int \!\! r \(F_{ad}\)_{r}d^{3}\vec{r}\right/ V,
\ee
where \[V=\frac{4\,\pi\,R_{s}^3}{3}\] is the volume of the spherical swarm.
For a linear restoring force of the form
\be
\(F_{ad}\)_{r}=-K\,r
\ee
where $K$ is a positive constant,
we get
\be
\bar{W}=-\frac{3}{5}K\,R_{s}^2.
\ee
In the case of gravitational interaction without adaptivity (Eq.~(\ref{New_grav})), the effective spring constant is
\be
K=\frac{4}{3}\pi\,\rho\, C,
\ee
and then we get a quadratic dependence on $R_s$:
\be
\bar{W}=-\frac{4}{5}\pi\,\rho\,C\,R_s^2.
\label{WaRegular}
\ee

The force with the adaptive correction is obtained by substituting Eq.~(\ref{total noise}) into Eq.~(\ref{normalized force}):
\be
F_{eff}(r)=-\frac{4\,C\,r^2}{3\,R_{ad}^2\,[2\,R_s\,r-(R_s^{2}-r^{2})\ln\(\frac{R_s-r}{R_s+r}\)]}.
\ee
Substitution into Eq.~(\ref{pot_sphe}) gives:
\be
\bar{W}=-\frac{4\,C\,R_s}{R_{ad}^{2}}  \int_{0}^{1}\frac{x^5\,dx}{2\,x+(x^2-1)\,\ln\(\frac{1-x}{1+x}\)} \sim -0.3\frac{C}{R_{ad}^{2}}\,R_s
\ee
where $x\equiv r/R_s$.

Thus, for adaptive gravity in the purely adaptive regime, the potential energy behaves as
\be
|\bar{W}| \propto R_s.
\label{Wadaptive}
\ee

\begin{figure}[h]
     \includegraphics[width=1.0\linewidth]{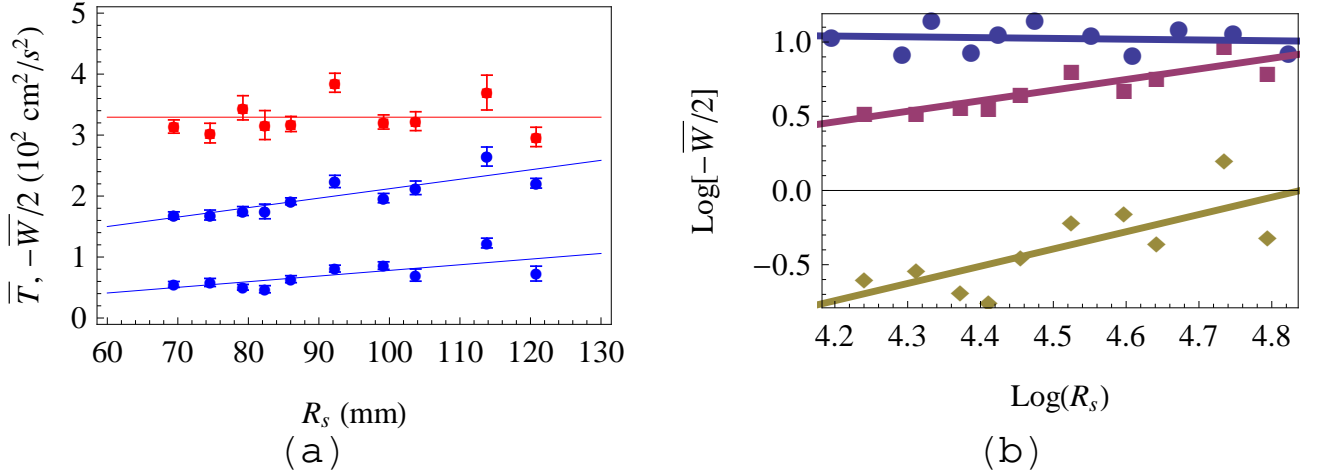}
\caption{(a) The mean kinetic $\bar{T}$ (red) and potential $-\bar{W}/2$ (blue, constrained for $r<R_s/2,R_s$ from bottom to top) energies as a function of $R_s$. (b) Log-Log plot for the potential energy in the center of the swarm with a linear fit, constrained for $r<R_s/2,R_s,\infty$ (yellow, purple and blue respectively). The power-law slopes are (yellow and purple respectively): $1.16\pm0.12, 0.71\pm0.05$.}
\label{EnergiesNearC.fig}
\end{figure}

We now compare this result to the potential energy contribution of the midges near the center of the swarm, where we expect to find the strongest effect of adaptivity. When we include all the midges of the swarm, the adaptivity is not significant since the midges with large radius ($r>R_{s}$) and low density dominate the contribution to the total potential energy. For this purpose we calculated the potential and kinetic energies of the same swarms (more than five midges) but this time the summation was carried out only up to an upper cutoff ($R_{s}$ and $R_s/2$). Near the center of the swarm the density is roughly constant and high, so that the adaptive calculation of Eqs.~(\ref{Wswarm})-(\ref{Wadaptive}) should apply. The results are presented in Fig.~\ref{EnergiesNearC.fig}. The mean kinetic energy is similar to the previous one (Fig.~\ref{Energiesfar.fig}), i.e. independent of $R_s$, except for some under-sampling of the fastest midges: $\langle\bar{T}\rangle_{(r<R_{s})}=(3.07\pm0.17)\cdot 10^2$ cm$^2$/s$^2$.
The potential energy is not constant and it is increasing as a function of $R_{s}$.
In Fig.~\ref{EnergiesNearC.fig}b we show that as the center of the swarm is approached (i.e. $r$ is constrained to smaller values), the increase of $\bar{W}$ with $R_s$ approaches a linear behavior, as we predict in Eq.~(\ref{Wadaptive}). Note that this is very different from the quadratic behavior for regular gravity (Eq.~(\ref{WaRegular})).
This observation therefore constitutes an additional independent and strong source of support for our adaptive-gravity form of the interactions within the midge swarm.

The virial relation also allows us to estimate the effect of the swarm size on the mean distance of closest approach between two midges, which was found to decrease for increasing swarm size. This relation is given in the Appendix (Eqs.~(\ref{feffmanynn})-(\ref{dnnAdaptive}), Fig.~\ref{rmin.fig}).

\subsection{Particle-based simulations.}

So far, we have demonstrated that the mean-field predictions of the AGM are in good agreement with the experimental results. To explore the AGM further, we performed molecular dynamics-type, agent-based simulations. To focus on the effects of the proposed adaptive-gravity interactions, we did not include in the simulation any explicit noise terms. Thus, the motion of each midge arises purely from their mutual interactions. However, to maintain numerical stability and cohesion of the swarm, it was necessary to augment the basic AGM (Eq.~(\ref{feffmany})) in three ways.

First, we added a short-range repulsion between midges. This repulsion prevents the fragmentation of the swarm into small groups (such as pairs or triplets) that can become effectively isolated from the rest of the swarm due to adaptivity-induced screening (Eq.~(\ref{feffmanynncond}), Fig.~\ref{swarmsSim}); additionally, we previously found experimental evidence for this kind of short-range repulsion in real swarms \cite{puckett2014}. With this repulsion, the effective force in Eq.~(\ref{feffmany}) becomes
\begin{equation}
\vec{F}^{i}_{eff}=C\sum_{j} \frac{\widehat{r}_{ij}}{r_{ij}^2} [1-2\exp(-(r_{ij}/R_r)^2]\left(\frac{R_{ad}^{-2}}{R_{ad}^{-2}+\sum_{k}{|\vec{r}_i-\vec{r}_k|^{-2}}}\right).
\label{F_midges}
\end{equation}

To prevent runaway midges and to be physically realistic, we also imposed a maximum midge velocity $v_{\text{max}}$. If the midge velocity exceeds this value, we re-scale it as
\begin{equation}
\vec{v}_{\text{new}}=\vec{v} v_{\text{max}}/v.
\label{v_max}
\end{equation}

And finally, we added an overall confining force to prevent the swarm from drifting in space, as discussed above. This force is significant only far from the swarm center, and is intended to model the attraction to ground-based visual features that localize natural swarms \cite{puckett2014determining}. It is given by
\begin{equation}
\vec{F}^{i}_{\text{marker}}=-\widehat{r_i}\frac{12}{R_{\text{marker}}}(r_i/R_{\text{marker}})^{11},
\label{F_spot}
\end{equation}
where the marker size is set by $R_{\text{marker}}=1.5R_{ad}$. We do not, however, impose any differences between the vertical direction and the in-plane directions, and so our simulated swarms are statistically isotropic in space.

The initial conditions in the simulations are as follows. The spatial coordinates for each midge
are expressed in spherical coordinates $r$, $\theta$, and $\varphi$, and are initially random variables uniformly chosen
between [0,$R_{\text{marker}}$], [0,$\pi$] and [0,$2\pi$], respectively. The initial velocities
are all zero. We then update the locations and velocities of each of the particles in time by solving the (Newton's) equations of motion of the particles (that is, $\dot{\vec{v}}^{i}=\vec{F}^{i}_{eff}+\vec{F}^{i}_{\text{marker}}$) using Runge-Kutta integration. Note that, unlike in real life and unlike in most models, the midges in these simulations do not have any intrinsic self-propelled motion; rather, their motion arises purely due to interactions, and can itself be viewed as an emergent property of the swarm.

\begin{figure}[tb]
\centering
\includegraphics[width=1\linewidth]{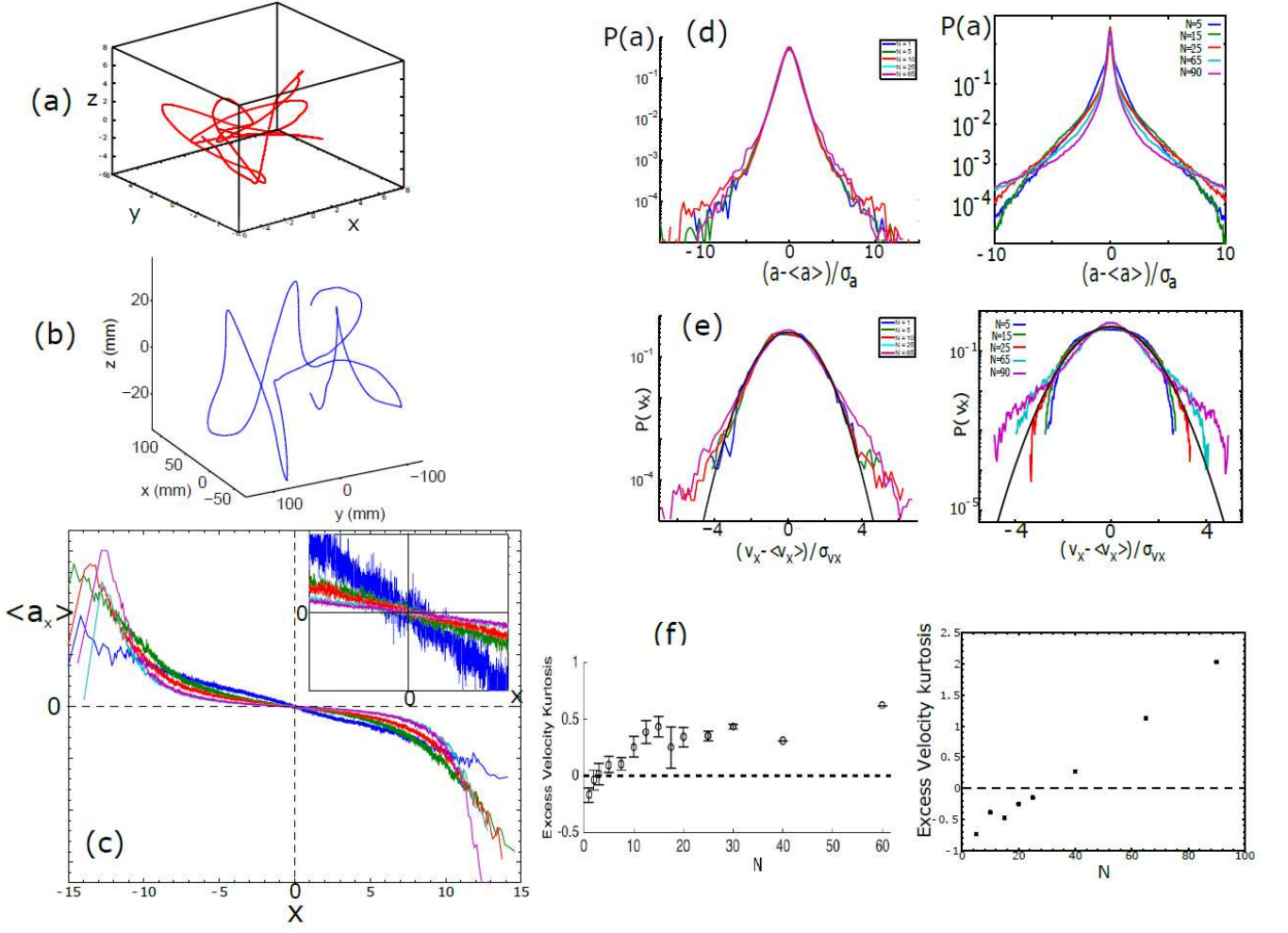}
\caption{\label{SimResults} Comparison of the adaptive-gravity simulations to experimental observations. (a)
A sample simulated trajectory from a swarm of $N=50$ midges, as compared with (b)
an observed trajectory in a swarm of $N=25$ midges \cite{puckett2014determining}. (c) Simulated mean acceleration along the $x$-axis
as a function of position for swarms of various sizes ($N=15,25,40,90$, from top to bottom). The top right inset shows the simulated linear behavior near
the swarm center, and the dependence of the slope on the swarm size (different colors) is similar to the observed behavior shown in Fig.~\ref{forceShape}a. (d) Observed acceleration distribution (left panel, \cite{kelley2013}), compared to the simulation results (right
panel). Both display highly non-Gaussian tails. (e) Observed $x$-component velocity
distribution (left panel, \cite{kelley2013}), compared to the simulation results (right panel). Both show that
small swarms have a roughly Gaussian velocity distribution, but large swarms develop heavy tails. This tendency is quantified in \textbf{f}, where the excess kurtosis is plotted as a function of
swarm size, and both observations (left panel, \cite{puckett2014determining}) and simulations (right panel) indicate a
negative excess kurtosis for small swarms and a positive kurtosis (due to the roughly exponential tails in the
distributions) for large swarms. All the simulations were carried out using $R_{ad}=10,R_r=0.3R_{ad},R_{marker}=1.5R_{ad},$ and $v_{max}=1$.}
\end{figure}

A sample trajectory of a single midge from a simulated swarm of $N=50$ midges, is shown in Fig.~\ref{SimResults}a, which qualitatively resembled an observed trajectory (Fig.~\ref{SimResults}b). The simulated mean acceleration of a midge towards the swarm center as a function of the distance from the center is shown in Fig.~\ref{SimResults}c. We recover the linear behavior near the swarm center, as expected, while the forces saturate near the swarm edge due to the Gaussian density profile of the swarm (Fig.~\ref{forceShape}b). The slopes near the center define the effective spring constants, and are found to decrease with the number of midges, just as they do in the experiments. In Fig.~\ref{SimResults}d, we plot the distribution of midge accelerations, and find that it displays the same qualitative features found in the experiments \cite{kelley2013}: the distributions are close to Gaussian for very small accelerations, but show heavy, exponential tails for large accelerations. And just as in the experiments, we find that these distributions are largely independent of the swarm size.

We also compared the velocity distributions from the simulations (Fig.~\ref{SimResults}e) as a function of the swarm size, and again found the same trend observed in the experiments \cite{kelley2013}: as the swarm size increases, the velocity distributions also develop a long exponential tail. This behavior can be quantified by calculating the excess kurtosis of the $x$-component velocity distribution as a function of the swarm size (Fig.~\ref{SimResults}f), which follows the same qualitative behavior seen in experiments \cite{puckett2014determining}. For very small swarm sizes, the excess kurtosis is slightly negative (meaning that the tails of the velocity distribution are slightly sub-Gaussian), and becomes positive for larger swarms.

Let us note that in these simulations we did not fit any parameters in an aim to reproduce the experimental observations quantitatively. Rather, we focus here on exploring the qualitative features that arise due to the adaptive-gravity interactions. It is quite satisfying that the distinctly non-Gaussian distributions of the accelerations (Fig.~\ref{SimResults}d) and of the velocity (Fig.~\ref{SimResults}e,f) already appear within our simple model, as well as the overall dependence of various features on swarm size. We anticipate that a more detailed model that includes, for example, the stochastic motion of an individual midge in isolation \cite{puckett2014determining} or the earth's gravitational field may be able to capture the mean-field behavior of the swarms quantitatively as well.

\section{Discussion}
We have presented here a new model of collective behavior in insect swarms that is based on the way that midges are thought to sense their environment, i.e. through acoustic signals. The AGM we have constructed introduces features that are not typically considered in models of collective motion, including long-range interactions and a sensitivity to the global properties of the group (through adaptivity). As we have shown, these features combine to produce group cohesion as a natural emergent property. Basic assumptions of the model, such as the precise relation between the received sound and the force produced by the midge await future direct experimentation, by, for example, studying external acoustic perturbations of swarms \cite{ni2015}. However, by comparing the predictions of the model with detailed statistical data extracted from real insect swarms measured in the laboratory, we have demonstrated that our model is able to capture not just the cohesion of swarms but also many of their many-body dynamical properties. The excellent agreement between the AGM and the observed behavior of both the spatial profile of the average forces within the swarm (Figs.~\ref{KvsRs},\ref{fig:v1}) and the virial relation (Fig.~\ref{EnergiesNearC.fig}), gives strong support to the model, and to its two main features: long-range (gravity-like) $1/r^2$ interactions and an adaptive response that renormalizes the effective forces according to the local noise amplitude.

To conclude, this model opens the door for further tests of the large-scale behavior and stability of swarms. Intriguingly, the success of the adaptive-gravity approach raises the question of whether some of the well known phenomena that occur in self-gravitating systems, such as the Jeans instability or gravitational collapse \cite{jeans1922motions,binney1982dynamics}, can occur in this biological system as well. This will be probed in future experiments, using acoustic perturbations of the swarm \cite{ni2015}. From the more general physics point of view, we introduce here a new class of ``adaptive'' interactions, which have novel physical features as well as apply to other biological systems -- for example in the context of chemical sensing, chemotaxis-driven interactions between swarming cells (see for example \cite{lammermann2013neutrophil}). This work opens a new class of physical systems with adaptive interactions, whose properties may be very different from their non-adaptive analogs.

\appendix
\begin{center}
      {\bf APPENDIX}
 \end{center}
 \renewcommand{\theequation}{A\arabic{equation}}
 \renewcommand{\thefigure}{A\arabic{figure}}

\setcounter{equation}{0}  
\setcounter{figure}{0}
\section{Adaptive-gravity potential between two midges}
\begin{figure}[h]
     \includegraphics[width=0.5\linewidth]{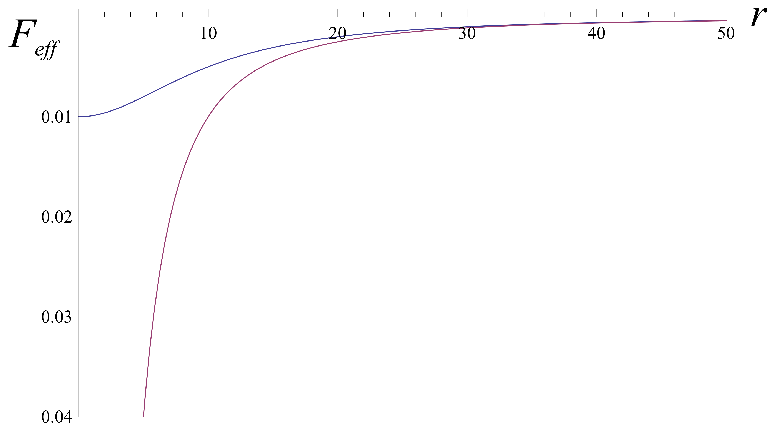}
\caption{Magnitude of the effective force $F_{eff}$ calculated according to Eqs.~(\ref{feff}) and~(\ref{adapt}) using $R_{ad}=10$ (blue line). The red line denotes pure gravity without adaptivity, i.e. for $\mathcal{A}=1$.}
\label{feff.fig}
\end{figure}

Let us consider two interacting midges. In this case the effective force (Eq.~(\ref{feffmany})) felt by one midge due to a second can be written as (taking $C=1$)
\begin{equation}
\vec{F}_{eff}=\hat{r}_{12}\frac{1}{|\vec{r}_1-\vec{r}_2|^2}\mathcal{A}(|\vec{r}_1-\vec{r}_2|) \label{feff}
\end{equation}
where $\hat{r}_{12}=(\vec{r}_1-\vec{r}_2)/|\vec{r}_1-\vec{r}_2|$ is the unit vector connecting the two midges and the adaptivity function is given by
\begin{equation}
\mathcal{A}(|\vec{r}_1-\vec{r}_2|)=\frac{R_{ad}^{-2}}{R_{ad}^{-2}+|\vec{r}_1-\vec{r}_2|^{-2}}. \label{adapt}
\end{equation}
$R_{ad}$ is a measure of the maximal distance between midges over which the adaptivity of the midge acoustic sensing can function. Beyond this distance, the sound reaching the midge is so weak that the hearing sensitivity (that is, the effective gain provided by the adaptivity) of the midge reaches its maximal value. See Fig.~\ref{feff.fig}. Note that as the midges come closer than $R_{ad}$, the effective force approaches a constant, while in pure gravity the force between two bodies increases without bound as they come together, leading to collapse. For $r_{12}\geq R_{ad}$ the effective force approaches the long-range gravity behavior of $1/r_{12}^2$.

\begin{figure}[h]
     \includegraphics[width=0.5\linewidth]{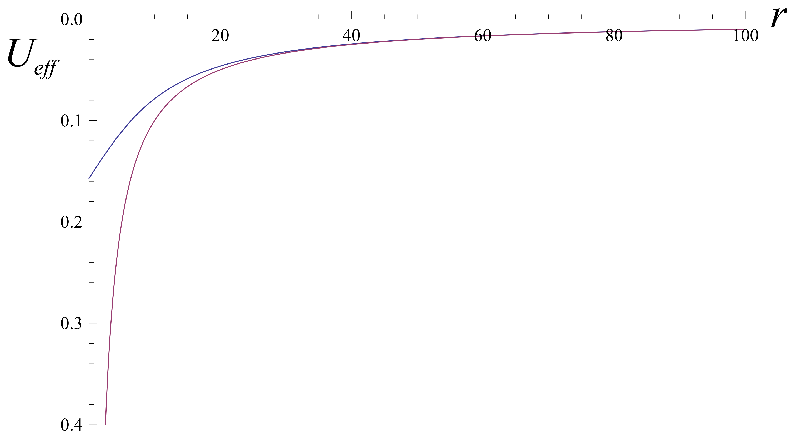}
\caption{Effective potential $U_{eff}$ calculated from Eq.~(\ref{ueff}) using $R_{ad}=10$ (blue line). The red line denotes pure gravity without adaptivity, i.e. for $\mathcal{A}=1$.}
\label{ueff.fig}
\end{figure}

We can integrate this force and calculate the effective midge-midge potential to be
\begin{equation}
U_{eff}(r_{12})=\frac{1}{R_{ad}}\left(\arctan{[r_{12}/R_{ad}]}-\pi/2\right).
\label{ueff}
\end{equation}
See Fig.~\ref{ueff.fig}. Unlike in gravity, where the potential energy diverges when the two particles overlap, due to adaptivity the potential approaches a finite value with a ``cusp'' shape, so that it is linear in $r_{12}$. For $r_{12}\geq R_{ad}$ the effective potential approaches the long-range gravity behavior of $1/r_{12}$.

\section{Explicit Calculation of the Effective Force on a Midge}

\begin{figure}[tb]
\centering
\includegraphics[width=0.25\linewidth]{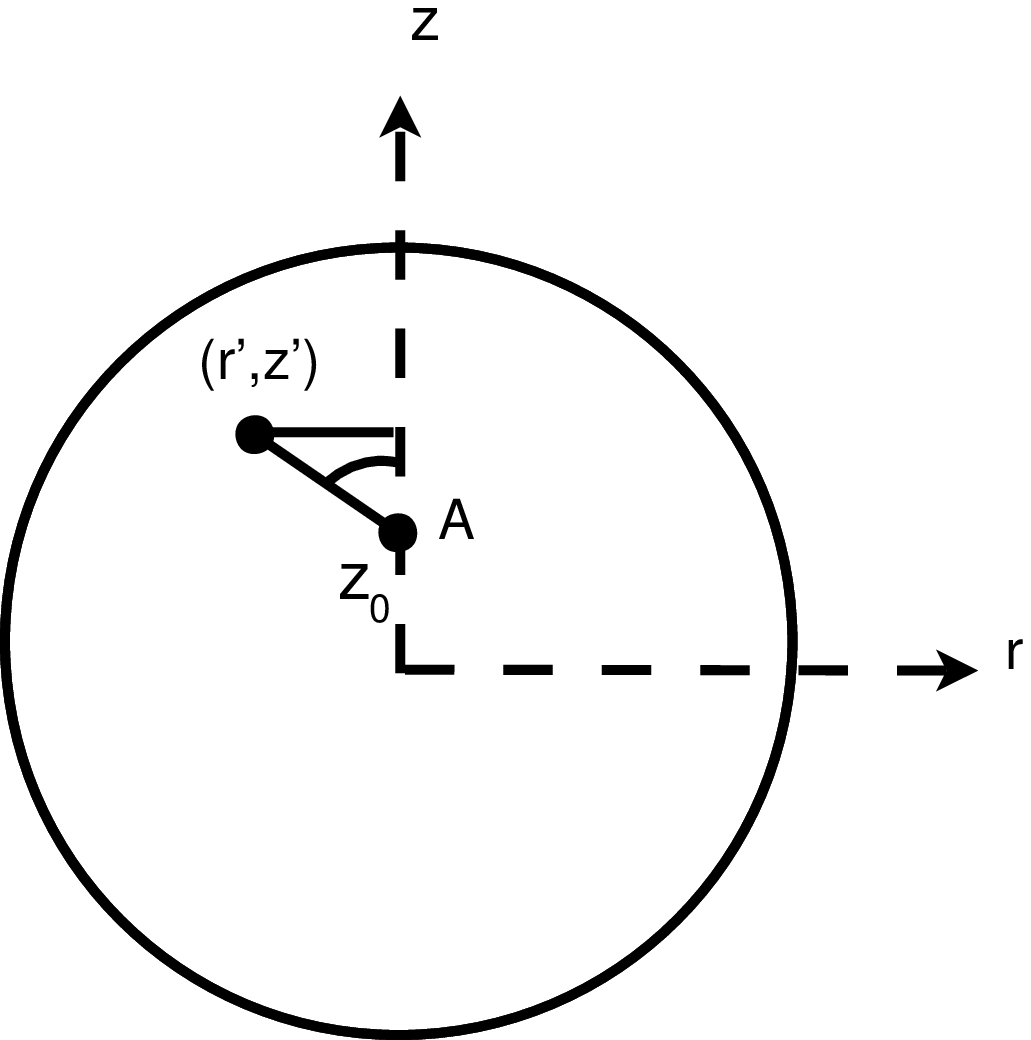}
\caption{\label{cylindricalcoordinates}  The cylindrical coordinates $(r,z)$ (and $\varphi$) that we use for the calculation of the effective gravitational field at a point A in a uniform-density spherical swarm.}
\end{figure}

Let us compute explicitly the adaptive-gravitational field at a point inside a spherical swarm with radius $R_s$ and uniform density $\rho$, according to Eq.~(\ref{normalized force}). We will use cylindrical coordinates $(r,z,\varphi)$ and calculate the field at  ($r=0$, $z=z_0$) without loss of generality (the point $A$ in Fig.~\ref{cylindricalcoordinates}). The symmetry of the problem implies that the field is along the $z$ axis. The contribution of a point at $(r',z')$ to the gravitational force at $(0,z_0)$ is
\[
\frac{1}{r'^2+(z'-z_0)^2},
\]
and the angle is
\[\cos\alpha=\frac{z'-z_0}{\sqrt{r'^2+(z'-z_0)^2}}.\]
Hence the gravitational force at $z_0$ is
\bea
F_{eff,g}(z_0)&=&2\,\pi\,C\,\rho\int_{-R_s}^{R_s}\!\!\!dz'\int_{0}^{\sqrt{R_{s}^2-z'^2}}\!\!\!\!\!r'dr'\,\frac{z'-z_0}{\left[r'^2+(z'-z_0)^2\right]^{\frac{3}{2}}} \nonumber\\
&=&-\frac{4\,\pi\,\rho\,C}{3}z_0,  \label{New_grav}
\eea
which agrees with the result obtained from Gauss's law in Eq.~(\ref{gauss}).
The normalization factor, which is given by the summation of the absolute values of the contributions to the point $(0,z_0)$, reads
\bea
N_{tot}(z_0)&=&2\,\pi\,\rho\,\int_{-R_s}^{R_s}dz'\int_{0}^{\sqrt{R_{s}^2-z'^2}}\frac{r'dr'}{r'^2+(z'-z_0)^2} \nonumber\\
&=&\pi\rho[2R_{s}-\frac{(R_s^{2}-z_0^{2})}{z_0}\ln\(\frac{R_s-z_0}{R_s+z_0}\)]. \label{total noise}
\eea
To leading order in $z_0$, we have
\be
N_{tot}(z_0)=4\pi\rho R_{s}+\mathcal{O}(z_0^2), \label{Ntot}
\ee
which confirms the $1/R_s$ leading behavior of the effective spring constant in the purely adaptive regime (Eq.~(\ref{normalized force})). We find that the profile of $F_{eff}(z)$ has the following behavior (Fig.~\ref{forceShape}a): The force is approximately linear in the distance out to at least half of the swarm size, and only grows faster than linear close to the edge. Note that we have assumed a uniform distribution here. Since the real distribution is close to Gaussian (see Fig.~(2) of \cite{kelley2013}) and decreases significantly when $r\sim R_{s}$, we do not expect to see such an increase in the force near the edges in real swarms.

\section{Limit of cylindrical swarm}

In the case of an elongated swarm in the form of a cylinder with radius $R$ and length $2\,L$, the gravitational force at $z_0$ is
\bea
F_{eff,g}(z_0)&=&2\,\pi\,C\,\rho\int_{-L}^{L}\!\!\!dz'\int_{0}^{R}\!\!\!\!\!r'dr'\,\frac{z'-z_0}{\left[r'^2+(z'-z_0)^2\right]^{\frac{3}{2}}} \nonumber\\
&=&-4\,\pi\,\rho\,C\,\(1-\frac{L}{\sqrt{L^2+R^2}}\)z_0+\mathcal{O}\(\frac{z_0^3}{L^3}\),
\eea
When we have a very long cylinder $L\gg R$,
\be
F_{eff,g}(z_0)\sim-2\,\pi\,C\,\rho\,\frac{R^2}{L^2}\, \label{cylin1}
z_0
\ee
The normalization factor in this case is
\bea
N_{tot}(z_0)&=&2\,\pi\,\rho\,\int_{-L}^{L}dz'\int_{0}^{R}\frac{r'dr'}{r'^2+(z'-z_0)^2} \nonumber\\
&=&4\,R\,\,\pi\,\rho\,\arctan\(\frac{L}{R}\)+2\,L\,\pi\,\rho\,\ln\(1+\frac{R^2}{L^2}\)+\mathcal{O}\(\frac{z_0}{L}\),
\eea
and for a long cylinder
\be
N_{tot}(z_0)\sim 2\,\pi^2\,\rho\,R. \label{cylin2}
\ee
From Eqs.~(\ref{cylin1}) and ~(\ref{cylin2}) we get the effective spring constant in the vertical direction as
\be
K_z \sim \frac{C\,R}{\pi\,L^2}
\ee
The radius of a long cylindrical swarm is
\be
R_s=\langle r \rangle=\frac{2\,\pi\,\int_{-L}^{L}dz\int_{0}^{R}\sqrt{r^2+z^2}\,r\,dr}{2\,\pi\,R^2\,L}\sim \frac{L}{2} + \mathcal{O}\(L^{-1}\) \label{size_cylinder}
\ee
Then in terms of the swarm radius
\be
K_z \propto R_s^{-2}.
\label{KzCylinder}
\ee

\section{Calculation of the effective spring constants in the ellipsoidal approximation}

Here we show how to derive Eqs.~(\ref{spring1})-(\ref{spring3}). We start from a computation of the gravitational potential inside an ellipsoid, and from it we will derive the effective ``spring constants.'' The derivation here follows Ref.~\footnote{Based on: Wei Cai, ``Appendix - Potential Field of a Uniformly Charged Ellipsoid,'' as part of the lecture notes to the course: ``Theory and Applications of Elasticity,'' Department of Mechanical Engineering, Stanford University, CA 94305-4040. http://micro.stanford.edu/$\sim$caiwei/me340/A\_Ellipsoid\_Potential.pdf}.

Let us start from an ellipsoid centered at the origin such that
\be
\frac{x^2}{a^2}+\frac{y^2}{b^2}+\frac{z^2}{c^2}=1, \label{MainEllipsoid}
\ee
whose semi-axes are $a$, $b$, and $c$, and where we assume $c<b<a$ without loss of generality. Let us now consider a family of ellipsoids
\be
\frac{x^2}{a^2}+\frac{y^2}{b^2}+\frac{z^2}{c^2}=u^2 \label{Shell}
\ee
for $0<u<1$.

Our goal is to find the gravitational potential of an ellipsoid~(\ref{MainEllipsoid}) with a uniform mass density $\rho$.
For this purpose, we will calculate the potential due to an ellipsoidal shell between $u$ and $u+du$ and then sum over all shells for $0<u<1$. It is convenient to introduce ellipsoidal coordinates $(\lambda,\mu,\nu)$ defined by
\bea
x^2=\frac{(a^2\,u^2+\lambda)(a^2\,u^2+\mu)(a^2\,u^2+\nu)}{u^4\,(a^2-b^2)(a^2-c^2)},\\
y^2=\frac{(b^2\,u^2+\lambda)(b^2\,u^2+\mu)(b^2\,u^2+\nu)}{u^4\,(b^2-a^2)(b^2-c^2)},\\
z^2=\frac{(c^2\,u^2+\lambda)(b^2\,u^2+\mu)(b^2\,u^2+\nu)}{u^4\,(c^2-a^2)(c^2-b^2)},
\eea
where the ellipsoidal coordinates $(\lambda,\mu,\nu)$ have the following ranges:
\be
\lambda>-c^2\,u^2>\mu>-b^2\,u^2>\nu>-a^2u^2.
\ee
Note that in the definition of ellipsoidal coordinates that we use, $\lambda,\mu,\nu$ depend on $u$ since we define them for the ellipsoidal shell~(\ref{Shell}).
Surfaces of constant $\lambda$ are ellipsoids:
\be
\frac{x^2}{a^2\,u^2+\lambda}+\frac{y^2}{b^2\,u^2+\lambda}+\frac{z^2}{c^2\,u^2+\lambda}=1. \label{lambda}
\ee
Thus, $\lambda$ is the equivalent of the radial coordinate for ellipsoids. The original shell surface is $\lambda(u)=0$ and $\lambda(u)>0$ covers the region outside the shell~(\ref{Shell}).
A natural assumption (based on the ellipsoidal symmetry) is that the potential depends only on $\lambda$ and not on $\mu$ or $\nu$. In this case, the Laplace equation outside the shell (for $\phi(\lambda,u)$, $\lambda>0$) is
\be
\frac{4\sqrt{\beta(\lambda,u)}}{(\lambda-\mu)(\lambda-\nu)}\frac{\partial}{\partial \lambda}\(\sqrt{\beta(\lambda,u)}\,\frac{\partial \phi}{\partial \lambda}\)=0,
\ee
where
\be
\beta(\lambda,u) \equiv (a^2\,u^2+\lambda)(b^2\,u^2+\lambda)(c^2\,u^2+\lambda).
\ee
The solution of this equation can be written as
\be
\phi(\lambda, u)=-\int_{\lambda(u)}^{\infty}\frac{u^2\,D\,ds}{\sqrt{\beta(s,u)}},
\ee
where $D$ is a constant that will be related to the surface mass density of the shell, and the $u^2$ factor is due to the fact that the total charge scales as $u^2$.
Using Gauss's law for the shell,
\be
 -\left. \nabla\phi \cdot \hat{n}_{\lambda}\right|_{\lambda=0}=4\,\pi\,\sigma,
\ee
we get
\be
-\left.\frac{1}{\sqrt{g_{\lambda\lambda}}}\frac{\partial \phi}{\partial\lambda}\right|_{\lambda=0}=4\,\pi\,\sigma,
\ee
where $g_{\lambda\lambda}$ comes from the metric in ellipsoidal coordinates:
\be
g_{\lambda\lambda}=\(\frac{\partial x}{\partial \lambda}\)^2+\(\frac{\partial y}{\partial \lambda}\)^2+\(\frac{\partial z}{\partial \lambda}\)^2=\frac{(\lambda-\mu)(\lambda-\nu)}{4\,\beta(\lambda,u)},
\ee
so that
\be
D=2\,\pi \sigma \sqrt{\mu\nu}. \label{C}
\ee
On the other hand, the surface mass density can be expressed as
\be
\sigma=\rho\,dh,
\ee
where $dh$ is the local thickness of the shell.
In order to find the thickness $dh$, let us start with a calculation of the distance from the origin to a point $(x,y,z)$ on the surface of the ellipsoid shell.
The normal to the point $(x,y,z)$ is
\be
\hat{n}=\(\frac{x}{a^2},\frac{y}{b^2},\frac{z}{c^2}\)/\sqrt{\frac{x^2}{a^4}+\frac{y^2}{b^4}+\frac{z^2}{c^4}},
\ee
 and the distance of the origin to a point $(x,y,z)$ on the surface of the ellipsoid shell is then
 \be
 h=(x,y,z)\cdot \hat{n}=u^2/\sqrt{\frac{x^2}{a^4}+\frac{y^2}{b^4}+\frac{z^2}{c^4}}=\frac{u\,a\,b\,c}{\sqrt{\mu\nu}}.
 \ee
 Therefore the local thickness is
 \be
 dh=du\frac{a\,b\,c}{\sqrt{\mu\nu}},
 \ee
which allows us to express the surface density $\sigma$ of the shell in terms of $\rho$ as
 \be
 \sigma=\rho \frac{a\,b\,c}{\sqrt{\mu\nu}} \, du.
 \ee
 Substituting into the expression for $D$ (Eq.~(\ref{C})), we get
 \be
 D=2\,\pi\,a\,b\,c\,\rho\,du.
 \ee
 The potential of such a shell is thus
 \be
\phi(\lambda, u)du=-2\,\pi\,a\,b\,c\,\rho\,u^2\,du\int_{\lambda(u)}^{\infty}\frac{ds}{\sqrt{\beta(s,u)}}.
\ee
Then the gravitational potential of an ellipsoid is obtained by integration over such shells,
\be
U(\lambda)=\int_{0}^{1}\phi(\lambda,u)du.
\ee
Let us change variables to $v\equiv\lambda/u^2$, $t\equiv s/u^2$, so that
\be
U(v)=-2\,\pi\,a\,b\,c\,\rho\int_{0}^{1}\,u\int_{v(u)}^{\infty}\frac{dt}{\sqrt{\beta(t)}}du,
\ee
where $\beta(t)\equiv (a^2+t)(b^2+t)(c^2+t)$.
Since we are interested in the force inside the ellipsoid, let us consider the potential field at a point $(x_0,y_0,z_0)$ inside the ellipsoid. It corresponds to a shell at $u_0$ ($0<u_0<1$):
 \be
\frac{x_0^2}{a^2}+\frac{y_0^2}{b^2}+\frac{z_0^2}{c^2}=u_0^2.
\ee
For shells inside this shell ($u<u_0$) the point $(x_0,y_0,z_0)$ is outside and therefore $v=v(u)$ is the lower bound of the integral. Shells outside this shell ($u>u_0$) contribute the same value as on the shell surface (since $u_0$ is inside the shells) and then the lower bound is $v=0$. Therefore
\be
U(v)=-2\,\pi\,a\,b\,c\,\rho\left[\int_0^{u_0}u\left(\int_{v(u)}^{\infty}\!\!\!\frac{dt}{\sqrt{\beta(t)}}\right)du +\left(\int_{u_0}^{1}\!\!\!u \,du\right)\left(\int_0^{\infty}\!\!\!\frac{dt}{\sqrt{\beta(t)}}\right)\right]. \label{1potent}
\ee
Integration by parts of the first integral gives
\be
\int_0^{u_0}\!\!\!u\left(\int_{v(u)}^{\infty}\!\!\!\frac{dt}{\sqrt{\beta(t)}}\right)du=\left[\frac{u^2}{2}\int_{v(u)}^{\infty}\!\!\!\frac{dt}{\sqrt{\beta(t)}}\right]^{u_0}_{0}+\int_{0}^{u_0}\!\!\!\frac{u^2}{2\,\sqrt{\beta(v)}}\frac{dv}{du}du.
\ee
Notice that $v(u_0)=0$ and $v(0)=\infty$. Hence,
\be
\int_0^{u_0}\!\!\!u\left(\int_{v(u)}^{\infty}\frac{dt}{\sqrt{\beta(t)}}\right)du=\frac{u_0^2}{2}\int_0^{\infty}\!\!\!\frac{dt}{\sqrt{\beta(t)}}-\frac{1}{2}\int_{0}^{\infty}\!\!\!\frac{u^2\,dv}{\sqrt{\beta(v)}}.
\ee
Together with the second integral in (\ref{1potent}) we get
\be
U(v)=-\pi\,a\,b\,c\,\rho\int_{0}^{\infty}\!\!\!\(\frac{1-u^2}{\sqrt{\beta(v)}}\)dv.
\ee
From (\ref{lambda}) we get the following relation between $u$ and $v$:
\be
u^2=\frac{x^2}{a^2+v}+\frac{y^2}{b^2+v}+\frac{z^2}{c^2+v}.
\ee
Then the potential is
\be
U(x,y,z)=-\pi\,a\,b\,c\,\rho\!\int_{0}^{\infty}\!\!\!\(1-\frac{x^2}{a^2+v}-\frac{y^2}{b^2+v}-\frac{z^2}{c^2+v}\)\frac{1}{\sqrt{\beta(v)}}dv \label{potential_ellipsoid}
\ee
for a point $(x,y,z)$ inside the ellipsoid.
Effectively, we have a harmonic potential with different ``spring constants'' in each direction. Taking the derivative of Eq.~(\ref{potential_ellipsoid}) with respect to the coordinates $x$,$y$ and $z$ gives us Eqs.~(\ref{spring1})-(\ref{spring3}), respectively.

To calculate the correction due to adaptivity to leading order, we have to write the expression for the total buzzing noise (\ref{total noise}) for the ellipsoidal case. The only change is in the limits of integration and the fact that there is no longer any cylindrical symmetry. Here we integrate over an ellipsoid and get
\bea\label{NtotEllipse}
N_{tot}(z_0)\!\!\!&=&\!\!\!2\,\rho\int_{0}^{2\,\pi}\!\!\!d\varphi\int_{0}^{c}\!\!\!dz'\int_{0}^{R_{el}}\!\!\!\frac{r'dr'}{r'^2+(z'-z_0)^2} \\
&=&\!\!\!2\,\sqrt{2}\,\rho\,a\,b\,c\int_{0}^{2\,\pi}\!\!\!d\varphi\frac{\arccoth\(\frac{\sqrt{2}\,a\,b}{\sqrt{Q(a,b,c,\varphi)}}\)}{\sqrt{Q(a,b,c,\varphi)}}+\mathcal{O}(z_0^2),\nonumber
\eea
where
\[
Q(a,b,c,\varphi)\equiv2\,a^2\,b^2-(a^2+b^2)c^2+(a^2-b^2)c^2\cos2\,\varphi,
\]
and
\[R_{el}^{2}\equiv\frac{1-\frac{z'^2}{c^2}}{\frac{\cos^2\varphi}{a^2}+\frac{\sin^2\varphi}{b^2}}.\]
It turns out that this expression is symmetric under the interchange of $a$, $b$ and $c$, and therefore there is no change to the ratios of the effective spring constants due to the asymmetries of the ellipsoid in the leading-order correction.

\section{Mean closest approach distance.}

\begin{figure}[h]
\includegraphics[width=0.5\linewidth]{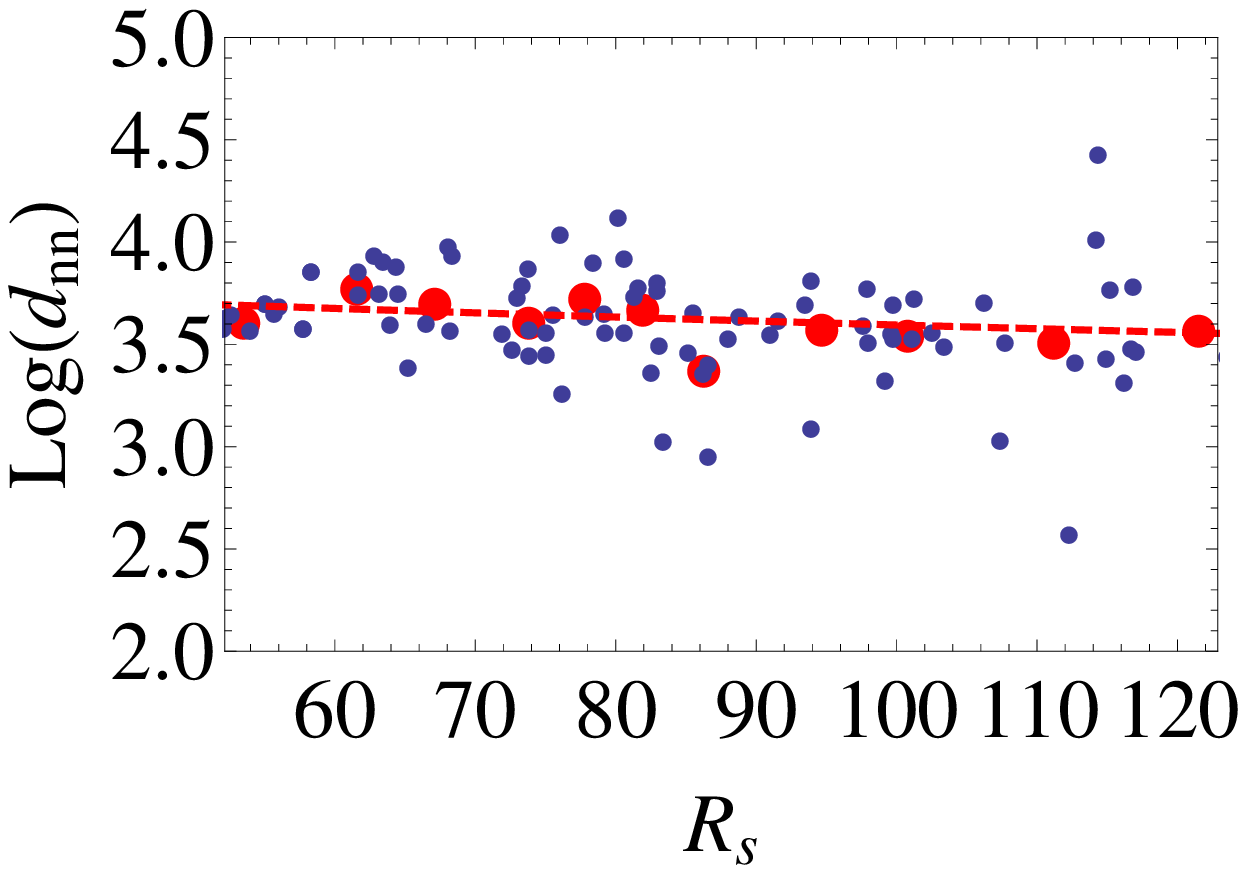}
\caption{Measured mean minimal midge separation $d_{nn}$ in different swarms (with at least four midges) \cite{puckett2014determining}, plotted against the mean swarm radius $R_s$ (blue circles: raw data, red circles: binned averages). We observe a decrease in the average minimal separation with increasing swarm size. The AGM predicts a shallow decrease, as indicated by the dashed line (Eq.~(\ref{dnnAdaptive})), which is in good agreement with the binned data (using the observed energies for $\bar{T},|\bar{W}|$ from Fig.~\ref{EnergiesNearC.fig}, and treating $\beta$ as the only fit parameter).}
\label{rmin.fig}
\end{figure}

It was observed that the mean nearest-neighbor distance $d_{nn}$ between midges decreases with increasing swarm size \cite{kelley2013,puckett2014determining,giardina2014prl} (Fig.~\ref{rmin.fig}). We now demonstrate that this decrease can be explained to arise from the adaptive nature of the interactions.
Due to the adaptive interactions (Eq.~(\ref{feffmany})), when two midges happen by chance to come very close to each other, they
may form a bound pair that is effectively screened from the rest of the swarm. In order to study the closest approach of midges in the swarm, let us consider the interaction of such a pair in the background provided by the rest. We assume that the separation between the two midges is small compared to their distance to the rest of the swarm, and thus that the interactions with the rest of the swarm will be negligible except for a contribution to the adaptivity factor in Eq.~(\ref{feffmany}), so that the effective force felt by one member of the pair is
\begin{equation}
\vec{F}_{eff,pair}\simeq C\hat{r}_{12}\frac{1}{|\vec{r}_1-\vec{r}_2|^2}\left(\frac{R_{ad}^{-2}}{R_{ad}^{-2}+I_{background}+|\vec{r}_1-\vec{r}_{2}|^{-2}}\right). \label{feffmanynn}
\end{equation}
$I_{background}$ is the sum over all the midges in the background. Near the center of the swarm this ``background noise'' takes, according to Eq.~(\ref{Ntot}), the form
\be
I_{background}=4\,\pi\,\rho\, R_{s},
\ee
where we assume spherical symmetry (radius $R_s$) and a constant density $\rho$ of the swarm.

Integrating this force we can calculate the effective two-body potential to be (see Eq.~(\ref{ueff}))
\be
U_{pair}=\frac{C}{\gamma\,R_{ad}}\(\arctan\(\frac{\gamma\,r_{pair}}{R_{ad}}\)-\frac{\pi}{2}\),
\ee
where $r_{pair}\equiv|\vec{r}_1-\vec{r}_2|$ and
\be
\gamma\equiv\sqrt{1+4\,\pi\,\rho\,R_{ad}^2\,R_{s}}.
\ee
We thus effectively have two-body motion under the influence of a mutual central force. Note that in the case of two bodies, additivity of the effective force is valid and as a result we can use all the conservation laws of a central force (namely energy and angular momentum).
 Therefore this two-body system can be reduced to an equivalent one-dimensional motion in the effective potential
\be
U_{eff,12}(r_{pair})=\frac{\beta}{r_{pair}^2}+\frac{C}{\gamma\,R_{ad}}\(\arctan\(\frac{\gamma\,r_{pair}}{R_{ad}}\)-\frac{\pi}{2}\),
\ee
where $\beta$ is a positive constant \footnote{In the case of gravity we take $\beta=\frac{l^2}{2\,\mu}$ where $l$ is the angular momentum and $\mu$ is the reduced mass of the system.} and the first term is the ``centrifugal barrier.'' Let us assume that the system of the two midges has an energy $E$. Then the minimal distance $d_{nn}$ is a turning point of the effective potential, given by the solution of the equation
\be
U_{eff,12}(d_{nn})=E.
\ee
For very short distances $r_{pair}\ll R_{ad}$, it is determined by the ``centrifugal barrier'', so that
\be
\frac{\beta}{d_{nn}^2}-\frac{\pi\,C}{2\,\gamma\,R_{ad}}=E.
\ee
For swarms where $ R_{s} <\sqrt{N} R_{ad}$, we can expand $d_{nn}$ to obtain
\be
d_{nn}=\(\frac{\beta}{E+\frac{\pi\,C}{2\,\gamma\,R_{ad}}}\)^{\frac{1}{2}}\simeq\sqrt{\frac{\beta}{E}}.
\label{dnnAdaptive}
\ee
For the central part of spherical swarms the mean energy of a midge depends on the swarm size (Eq.~(\ref{Wadaptive})), as $E=\bar{T}+|\bar{W}|$, and $|\bar{W}|\propto R_s$ (Fig.~\ref{EnergiesNearC.fig}). We therefore expect the closest approach distance to decrease with increasing swarm size.
The experimental data shown in Fig.~\ref{rmin.fig} for the center of the swarms (where our adaptivity calculation is applicable), indicates a very weak decrease with swarm size, consistent with this prediction.

\section{Tendency for swarm fragmentation due to the adaptive ``screening''}

\begin{figure}[tb]
\centering
\includegraphics[width=0.5\linewidth]{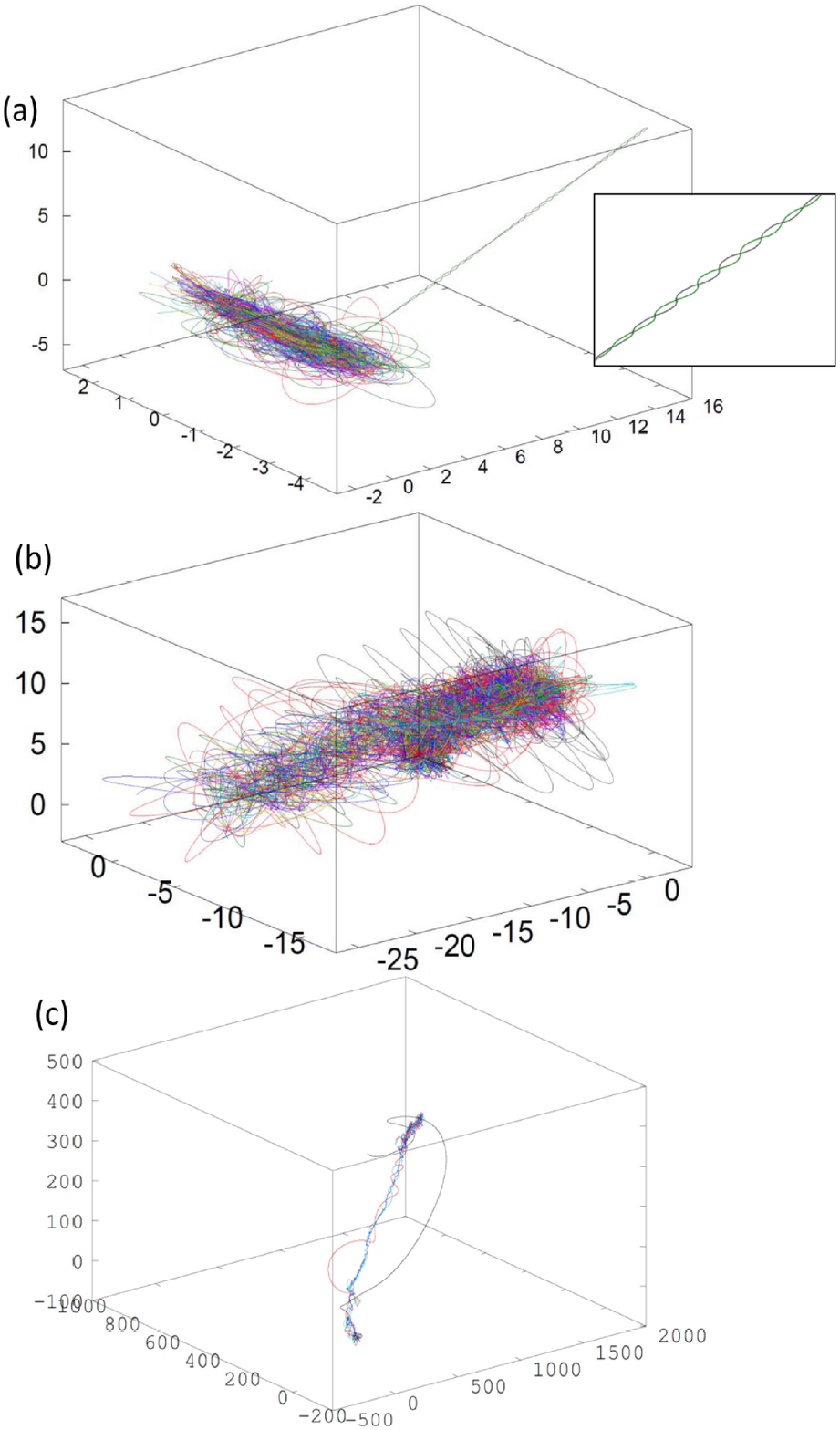}
\caption{(a) Traces of a simulated swarm with $N=50$ midges, demonstrating that without close-range repulsion there is a tendency to form very tight pairs that leave the swarm. (b) As in (a) but with short-range repulsion that prevents the formation of bound pairs, but no ``marker'' potential. (c) Traces of simulated trajectories of a swarm with $N=5$ midges, demonstrating the need for the ``marker'' potential and short-range repulsion to prevent midges from leaving the swarm.}
\label{swarmsSim}
\end{figure}

As discussed above, two nearby midges may form a bound pair due to screening by adaptivity. When the pair of midges are very close to each other, we can have that
\begin{equation}
\frac{1}{|\vec{r}_i-\vec{r}_{nn}|^2}\gg \frac{1}{R_{ad}^2}+I_{background} \label{feffmanynncond}
\end{equation}
which means that the interactions with all the other midges is reduced immensely by the huge factor of Eq.~(\ref{feffmany}) in the denominator. The resulting behavior is shown in Fig.~\ref{swarmsSim}. Including the ``marker'' potential and short-range repulsion (Eqs.~(\ref{F_midges})-(\ref{F_spot})) prevents midges from leaving the swarm (Fig.~\ref{swarmsSim}b,c), while otherwise pairs may form and shoot out of the swarm (Fig.~\ref{swarmsSim}a,b). Occasionally, larger number of midges (such as triplets) also form such fragments.

\bibliography{acoustics_1}

\textbf{Acknowledgements.}
N.S.G. and D.G. thank Sam Safran for useful discussion. J.G.P., R.N., and N.T.O. acknowledge support from the US Army Research Office under grant W911NF-13-1-0426.
N.S.G. gratefully acknowledges funding from the ISF (Grant
No. 580/12). This research is made possible in part by the
generosity of the Harold Perlman Family.

\end{document}